\documentclass[longauth]{aa}

\usepackage{graphicx}
\usepackage{txfonts}
\usepackage{booktabs}
\usepackage[flushleft]{threeparttable}
\usepackage[table]{xcolor}
\usepackage{amsmath}
\usepackage{array}
\usepackage{float}
\definecolor{cyan}{rgb}{0.88,1.0,1.0}
\definecolor{yellow}{rgb}{1.0,1.0,0.0}
\usepackage{multirow}
\usepackage{multicol}
\usepackage{blindtext}
\usepackage{subcaption}
\usepackage{longtable}
\begin{document}

   \title{The \textit{Gaia}-ESO Survey: a new approach to chemically characterising young open clusters\thanks{Based on observations collected with the FLAMES instrument at VLT/UT2 telescope
(Paranal Observatory, ESO, Chile), for the Gaia- ESO Large Public Spectroscopic Survey (188.B-3002, 193.B-0936).}$^,$\thanks{The full Table 2 is only available in electronic form at the CDS via anonymous ftp to cdsarc.u-strasbg.fr (130.79.128.5) or via http://cdsweb.u-strasbg.fr/cgi-bin/qcat?J/A+A/}}

   \subtitle{I. Stellar parameters, and iron-peak, $\alpha$-, and proton-capture elements}

   \author{M. Baratella\inst{1}
   \and V. D'Orazi\inst{2,3} 
   \and G. Carraro\inst{1} 
   \and S. Desidera\inst{2}
   \and S. Randich\inst{4} 
   \and L. Magrini\inst{4}
   \and V. Adibekyan\inst{5}
   \and R. Smiljanic\inst{6}
   \and L. Spina\inst{3}
   \and M. Tsantaki\inst{4}
   \and G. Tautvai{\v s}ien{\. e}\inst{7}
   \and S. G. Sousa\inst{5}
   \and P. Jofr\'e\inst{19}
   \and F. M. Jim\'{e}nez-Esteban\inst{8}
   \and E. Delgado-Mena\inst{5}
   \and S. Martell\inst{22,23}
   \and M. Van der Swaelmen\inst{4}
   \and V. Roccatagliata\inst{18,4}
   \and G. Gilmore\inst{14}
   \and E. J. Alfaro\inst{9}
   \and A. Bayo\inst{10,11}
   \and T. Bensby\inst{12} 
   \and A. Bragaglia\inst{13}
   \and E. Franciosini\inst{4}
   \and A. Gonneau\inst{14}
   \and U. Heiter\inst{15}
   \and A. Hourihane\inst{14}
   \and R.~D. Jeffries\inst{20}
   \and S.~E. Koposov\inst{21}
   \and L. Morbidelli\inst{4}
   \and L. Prisinzano\inst{16}
   \and G. Sacco\inst{4}
   \and L. Sbordone\inst{17}
   \and C. Worley\inst{14}
   \and S. Zaggia\inst{2}
   \and J. Lewis$^{\dagger}$
          }

 \institute{Dipartimento di Fisica e Astronomia {\it Galileo Galilei}, Vicolo Osservatorio 3, I-35122, Padova, Italy\\
\email{martina.baratella.1@phd.unipd.it}
    \and 
    INAF -- Osservatorio Astronomico di Padova, vicolo dell'Osservatorio 5, 35122, Padova, Italy
  \and
  Monash Centre for Astrophysics (MoCA), Monash University, School of Physics and Astronomy, Clayton, VIC 3800, Melbourne, Australia
  \and
  INAF -- Osservatorio Astrofisico di Arcetri, Largo E. Fermi 5, 
  50125, Firenze, Italy
  \and 
  Instituto de Astrofísica e Ciências do Espaço, Universidade do Porto, CAUP, Rua das Estrelas, 4150-762 Porto, Portugal
  \and
  Nicolaus Copernicus Astronomical Center, Polish Academy of Sciences, ul. Bartycka 18, 00-716, Warsaw, Poland
  \and 
  Institute of Theoretical Physics and Astronomy, Vilnius University, Sauletekio av. 3, 10257, Vilnius, Lithuania
  \and
  Departmento de Astrof\'{\i}sica, Centro de Astrobiolog\'{\i}a (INTA-CSIC), ESAC Campus, Camino Bajo del Castillo s/n, E-28692 Villanueva de la Cañada, Madrid, Spain
  \and
  Instituto de Astrof\'{i}sica de Andaluc\'{i}a-CSIC, Apdo. 3004, 18080, Granada, Spain
  \and
  Instituto de F\'isica y Astronom\'ia, Universidad de Valparaiso, Gran
Breta\~na 1111, Valpara\'iso
  \and
  N\'ucleo Milenio de Formaci\'on Planetaria, NPF, Universidad de
Valpara\'iso
  \and
  Lund Observatory, Department of Astronomy and Theoretical Physics, Box 43, SE-221 00 Lund, Sweden
  \and 
  INAF -- Osservatorio di Astrofisica e Scienza dello Spazio di Bologna, via Gobetti 93/3, 40129, Bologna, Italy
  \and
  Institute of Astronomy, University of Cambridge, Madingley Road, Cambridge CB3 0HA, United Kingdom
  \and
  Observational Astrophysics, Department of Physics and Astronomy, Uppsala University, Box 516, 75120 Uppsala, Sweden
  \and
  INAF - Osservatorio Astronomico di Palermo, Piazza del Parlamento 1, 90134, Palermo, Italy
  \and
  European Southern Observatory, Alonso de Cordova 3107 Vitacura, Santiago de Chile, Chile
  \and
  Dipartimento di Fisica “Enrico Fermi”, Universita’ di Pisa, Largo Pontecorvo 3, 56127 Pisa, Italy
  \and
  N\'ucleo de Astronom\'{i}a, Facultad de Ingenier\'{i}a, Universidad Diego Portales, Av. Ej\'ercito 441, Santiago, Chile
  \and
  Astrophysics Group, Keele University, Keele, Staffordshire ST5 5BG, United Kingdom
  \and
  McWilliams Center for Cosmology, Department of Physics, Carnegie Mellon University, 5000 Forbes Avenue, Pittsburgh, PA 15213, USA
  \and 
  School of Physics, University of New South Wales, Sydney, NSW 2052, Australia
  \and
  Centre of Excellence for All-Sky Astrophysics in Three Dimensions (ASTRO 3D), Australia
  \\ }

   \date{Received ??; accepted ??}


  \abstract
 {Open clusters are recognised as excellent tracers of Galactic thin-disc properties. At variance with intermediate-age and old open clusters, for which a significant number of studies is now available, clusters younger than $\lesssim$ 150 Myr have been mostly overlooked in terms of their chemical composition until recently (with few exceptions). On the other hand, previous investigations seem to indicate an anomalous behaviour of young clusters, which includes (but is not limited to) slightly sub-solar iron (Fe) abundances and extreme, unexpectedly high barium (Ba) enhancements.}
 {In a series of papers, we plan to expand our understanding of this topic and investigate whether these chemical peculiarities are instead related to abundance analysis techniques.}
 {We present a new determination of the atmospheric parameters for 23 dwarf stars observed by the \textit{Gaia}-ESO survey in five young open clusters ($\tau <$150\,Myr) and one star-forming region (NGC\,2264). We exploit a new method based on titanium (Ti) lines to derive the spectroscopic surface gravity, and most importantly, the microturbulence parameter. A combination of Ti and Fe lines is used to obtain effective temperatures. We also infer the abundances of Fe\,{\sc i}, Fe\,{\sc ii}, Ti\,{\sc i}, Ti\,{\sc ii}, Na\,{\sc i}, Mg\,{\sc i}, Al\,{\sc i}, Si\,{\sc i}, Ca\,{\sc i}, Cr\,{\sc i,} and Ni\,{\sc i}.}
{Our findings are in fair agreement with \textit{Gaia}-ESO iDR5 results for effective temperatures and surface gravities, but suggest that for very young stars, the microturbulence parameter is over-estimated when Fe lines are employed. This affects the derived chemical composition and causes the metal content of very young clusters to be under-estimated. }
{Our clusters display a metallicity [Fe/H] between +0.04$\pm$0.01 and +0.12$\pm$0.02; they are not more metal poor than the Sun. Although based on a relatively small sample size, our explorative study suggests that we may not need to call for ad hoc explanations to reconcile the chemical composition of young open clusters with Galactic chemical evolution models.}

\maketitle
%

\section{Introduction}

Open clusters (OCs) are among the most efficient objects with which the chemical properties and the evolution of the Galactic disc can be probed. These systems represent the concept of single stellar population well, that is, a group of coeval, (initially) chemically homogeneous stars. OCs cover a wide range in metallicity (between $-$0.3\,dex and +0.4\,dex), and most importantly, are almost ubiquitous in the Galactic disc. They therefore allow us to investigate a number of aspects, including stellar evolution and nucleosynthesis models, along with the radial and azimutal metallicity gradients (e.g. \citealt{1995Friel,2015donati,neto,2016jaco,2016reddy,2017mag}).\\
\indent
In recent years, the number of studies on OCs and their characterisation has enormously increased through the data from several spectroscopic surveys (e.g. the APO Galactic Evolution Expreiment, APOGEE, \citealt{cunha}, \citealt{2018donor}, \citealt{2019carrera}; and the Open Clusters Chemical Abundances from Spanish Observatories, OCCASO, \citealt{casamiquela} and references therein). However, while the number of studies on intermediate-age and old OCs ($\tau \gtrsim 600$ Myr) is conspicuous, less attention has been payed to the chemical composition of young OCs (YOCs, ages younger than $\sim$ 150 Myr) and star-forming regions (SFRs). The \textit{Gaia}-ESO public spectroscopic Survey \citep{gaia,randich} places special emphasis on observations of OCs covering the age range from a few million to several billion years, analysed in a homogeneous way. It therefore offers the opportunity of deriving the metallicity and abundances of a significant number of young objects \citep{2014spinaB,2014spinaA}.
The study of such systems is indeed very important to shed light on different topics, such as  the metallicity distribution in the solar neighbourhood, the present-day metallicity distribution in the Galactic disc \citep{Spina17}, the connection between the occurrence of giant planets and metallicity of the host stars (e.g. \citealt{2003santos}), and the behaviour of heavy-element (\textup{slow} n-capture process) abundances (see e.g. \citealt{2017dor} and references therein).\\
\indent
Somewhat at odds with what is expected from the Galactic chemical evolution models, several authors found that no YOCs or SFRs with super-solar metallicities exist \citep{james,2008santos,2011dor, 2011biazzoA,2011biazzoB,2014spinaB,2014spinaA,Spina17,2019origlia}. All the young populations in the solar neighbourhood seem to be slightly metal poor with respect to the Sun.
The presence of these systems might be explained with a complex combination of star formation, gas inflows and outflows, radial migration \citep{minchev}, or a different composition of the parental molecular cloud \citep{Spina17}.\\
\indent
It is well established that the frequency of giant planets is higher  around metal-rich stars, and this is predicted by planet formation models (core accretion; \citealt{pollack96}; and the recent tidal downsizing models; \citealt{2017nayakshin}). The lack of metal-rich YOCs and SFRs could influence this relation, and the question arises whether it is less probable to find giant planets in these systems.
There is growing observational evidence that in some cases, the planetary companion might be more metal rich than the host star (\citealt{vigan}, \citealt{samland}), as predicted by the core accretion paradigm. However, the uncertainty on the metallicity of the young stars hosting planets, with typical uncertainties of  $\sim$0.2 dex (e.g. \citealt{james}), prevents us from placing strong constraints on this aspect. 
This topic clearly deserves deeper investigation.\\
\indent
There are reasons to think that the sub-solar metallicity of young stars is not intrinsic, but related to analytical problems. Young stars have active chromospheres that may affect line formation and thus the derived stellar parameters and abundances. They also have intense magnetic fields \citep{2016folsom} that might be related to different phenomena and mechanisms. For instance, \cite{james} and \cite{2008santos} reported extremely high values of microturbulence parameters ($\xi$)  (up to $\sim$ 2.5\,km\,s$^{-1}$), along with a substantial star-to-star variation (that is apparently unrelated to the other stellar parameters: two stars in IC\,4665 analysed by Gaia-ESO have a difference in temperature of +90\,dex according to the iDR5 results and a difference in $\log g$ of +0.01\,dex, but a difference in $\xi$ of +0.70\,kms$^{-1}$). A similar behaviour has also been reported \cite{viana}, who analysed stars in 11 associations containing young stars, and found $\xi$ values up to $\sim$ 2.6\,km\,s$^{-1}$. Moreover, the authors reported a weak trend of $\xi$ with the effective temperature ($T\rm{_{eff}}$), with higher values at lower $T\rm{_{eff}}$. 

Recently, \cite{reddy} have investigated a possible correlation between Fe~{\sc i} abundances and the line formation optical depth. According to their results, the Fe~{\sc i} lines that form in the upper layers of the photosphere in young active stars provide larger abundances than those forming in the lower layers, probably because of the active chromosphere. This  trend may also affect the value of $\xi$ when it is derived by removing the slope between individual Fe\,{\sc i} line abundances and their reduced equivalent widths (REWs).  $\xi$ is a free parameter introduced to account for the difference between the observed equivalent widths (EWs) of moderate to strong lines and those predicted by models based on thermal and damping broadening alone. Weaker lines are almost independent of this parameter; it is therefore calculated by forcing weak and strong lines to be in agreement.  If strong lines (typically forming in the upper photospheric layers) yield anomalously larger abundances than weaker lines,  $\xi$ needs to be increase in order to remove the slope.  \cite{2019galarza} recently reported on the effects of stellar activity on the stellar parameters, but their analysis involved a star that is much older ($\tau\sim$400-600\,Myr) than those in our sample. \\
\indent
Another important aspect is the behaviour of the s-process elements. \cite{2009dor} found that the [Ba/Fe] ratio is positively correlated with OC ages. The standard analysis of the Ba\,{\sc ii} 5853\,\AA\,and 6496\,\AA \,lines gave values up to +0.60\,dex in clusters younger than 50\,Myr, but solar values in older clusters (age $\gtrsim$ 1-2\,Gyr). The increasing trend as a function of the age has been confirmed by other authors \citep{maiorca,jacobson,mishe,reddy,magrini,2019delgadomena}, and it has been interpreted by \cite{2012maiorca} as due to the recent production by nucleosynthesis in asymptotic giant branch (AGB) stars of low mass. This can explain mild enhancements (up to 0.1-0.2\,dex), but not the much higher values measured in very young clusters, however. Of the different explanations, the sensitivity of the Ba\,{\sc ii} 5853\,\AA\, line to the $\xi$ parameter value seems to be the most promising \citep{2015red}. The Ba\,{\sc ii} 5853\,\AA\, forms in the upper layers of the photosphere, where the effects of the active chromosphere are stronger. All these pieces of evidence support our hypothesis that the standard chemical analysis in very young stars ($\tau$ < 150\,Myr) might lead to misinterpreted results. We will investigate the behaviour of s-process elements and their time evolution in a companion paper (Baratella et al., in preparation).\\
\indent

In this first paper of a series focused on the chemical characterisation of very young stars, we propose a new approach to perform the chemical analysis of these young objects. 
We analyse \textit{Gaia}-ESO fifth internal data release (iDR5) spectra of 23 stars observed in five YOCs plus one SFR by the \textit{Gaia}-ESO Survey. The dataset is described in Sec.\ref{S2}. In Sec.\ref{S3} we derive the input values of the atmospheric parameters from photometry. We determine the stellar parameters by employing a new method (Sec.\ref{S4}) that is almost entirely based on the use of \textit{\textup{titanium}} lines. We also derive abundances for different $\alpha$-, proton-capture, and iron-peak elements: Fe\,{\sc i}, Fe\,{\sc ii}, Ti\,{\sc i}, Ti\,{\sc ii}, Na\,{\sc i}, Mg\,{\sc i}, Al\,{\sc i}, Si\,{\sc i}, Ca\,{\sc i}, Cr\,{\sc i,} and Ni\,{\sc i}. In Sec.\ref{S5} and \ref{S6} we present our results and discuss the scientific implications.

\begin{table*}
\caption{Basic information of the SFR and YOCs investigated in this work.}
\label{tabinfo}
\centering
\begin{threeparttable}
\begin{tabular}{lcccccccccr}
\toprule
Cluster & RA & Dec & Age$^{\ast}$ & Distance & $R\rm{_{Gal}^{\ast}}$  &  $E(B-V)$ & Ref.\\
&\footnotesize{(J2000)} & \footnotesize{(J2000)} & \footnotesize{(Myr)}&\footnotesize{(pc)}&\footnotesize{(kpc)}&\footnotesize{(mag)}\\
\midrule
IC\,2391 &08 40 32.00 & $-$53 02 00.00 & 50$\pm$30 & 151$\pm$2& 8.00$\pm$0.01 & <0.05  & 1,2\\
IC\,2602 & 10 42 58.00 & $-$64 24 00.00 & 30$\pm$20 & 152$\pm$3  & 7.95$\pm$0.01 & 0.02-0.04 &3 \\
IC\,4665 & 17 46 18.00 & +05 43 00.00 & 40$\pm$10 & 345$\pm$12 & 7.72$\pm$0.01& 0.16-0.19 & 4 \\
NGC\,2264 & 06 40 58.00 &  +09 53 42.00 & 3-5$\pm$4$^{\ast \ast}$ & 723$\pm$57  & 8.66$\pm$0.13& 0.075 & 5,6  \\
NGC\,2516 & 07 58 04.00 & $-$60 45 12.00 & 130$\pm$60 & 409$\pm$18  & 7.98$\pm$0.01& 0.11 & 7 \\
NGC\,2547 & 08 10 25.70 & $-$49 10 03.00 & 50$\pm$20 & 387$\pm$15  & 8.05$\pm$0.01& 0.12  & 8\\
\hline
\end{tabular}
\tablebib{
1) \cite{1999bar}; 2) \cite{2004bar}; 3) \cite{2009van}; 4) \cite{car}; 5) \cite{2004sung}; 6) \cite{turner}; 7) \cite{18bai}; 8) \cite{nay}
}
\begin{tablenotes}
\small
\item Notes: The clusters are sorted by name. The distances (fifth column) are from \cite{2018cantat}. The asterisk indicates ages and Galactocentric distances ($R\rm{_{Gal}}$) from \cite{neto}. We adopted $R_{\rm{Gal,}\odot}$=8.00\,kpc. The double asterisk indicates the age value from \cite{venuti}.
\end{tablenotes}
\end{threeparttable}
\end{table*}

\section{Dataset}\label{S2}

We analyse high-resolution (R$\sim$47000) spectra of 23 solar-type dwarf stars (with spectral type from F9-K1) in five YOCs and one SFR, observed in the \textit{Gaia}-ESO Survey. The selected targets are IC\,2391, IC\,2602, IC\,4665, NGC\,2264, NGC\,2516, and NGC\,2547, all with ages younger than 150\,Myr. We have chosen these clusters because no observational studies have been carried out in the framework of heavy-element abundances, with the exception of IC\,2391 and IC\,2602, which we used as calibrators of our abundance scale. Some information on these objects is reported in Table \ref{tabinfo}.

The spectra of the target stars were acquired with the high-resolution Fiber Large Array Mulit-Element Spectrograph and the Ultraviolet and Visual Echelle Spectrograph (FLAMES-UVES) \citep{uves} and have been reduced by the \textit{Gaia}-ESO consortium in a homogeneous way. The data reduction of UVES spectra was carried out using the FLAMES-UVES ESO public pipeline \citep{modi,sacco}.

Different Working Groups (WGs) contribute to the spectrum analysis: for the stars considered here, the analysis was performed by WG11 and WG12. The details of the procedures are described in \cite{smi} and \cite{lanza}. The recommended parameters produced by this analysis are reported in the iDR5 catalogue and are used as comparison for the results we obtained with our new approach.

The UVES observations were performed with the 580 nm setup for F-, G-, and K-type stars, with the spectra covering the 4800-6800\,\AA\, wavelength range. In particular, this spectral range contains the 6708\,\AA\, line of $^{7}$Li, which is an important diagnostic of stellar age. We did not consider the GIRAFFE spectra because their spectral range is limited and they have a lower resolution. Of all the available spectra, we selected only stars with rotational velocities $v\,{\rm sin}\, i < 20$~km\, s$^{-1}$ to avoid significant or heavy line blending, with signal-to-noise ratios (S/N) higher than 50 and effective temperatures $T\rm{_{eff}}\gtrsim5200$\,K to avoid over-ionisation and/or over-excitation effects \citep{schu}. All selected stars are confirmed members of corresponding clusters through radial velocities (RVs) and the strength of the $^{7}$Li absorption line at 6708\AA, according to the \textit{Gaia}-ESO iDR5 measurements.

\section{Input estimates of the atmospheric parameters}\label{S3}

$T\rm{_{eff}}$ estimates were obtained using  photometry from the Two Micron All-Sky Survey (2MASS) \citep{2003cutri} with the calibrated relation by \cite{casagrande}, valid for $(J-K)$ de-reddened colours in the range $0.07<(J-K)_0 <0.80$\,mag. We used this photometry because it provides homogeneous data for the stars in this study. 

We used the classical equation for the surface gravity ($\log g$), 
\begin{equation}
\begin{split}
\log g= &\log g_{\odot}+\log\left(\frac{m_{\star}}{m_{\odot}}\right)+4\cdot \log\left(\frac{T_{\rm eff}}{T_{\rm eff,\odot}}\right)\\
&+0.4\cdot(M_K+BC_K-M_{BC,\odot}),
\end{split}
\end{equation}

where $T_{\rm eff,\odot}$=5771\,K, $\log g_{\odot}$=4.44\,dex, and M$_{BC, \odot}$=4.74. $T\rm{_{eff}}$ is the $T(J-K)$ estimate and $\rm{M_K}$ is the absolute magnitude in K band, calculated with the distance estimates reported in Table \ref{tabinfo}. $\rm{BC_K}$ is the bolometric correction in K band, calculated as in \cite{masana}.
The values of m$_{\star}$ were estimated using the Padova suite of isochrones \citep{2017mar}. From these, we infer the mass to be equal to 1\,$\rm{M_{\odot}}$ for the five YOCs, while for the stars in the SFR, we infer m$_{\star}$=2-3\,$\rm{M_{\odot}}$.

The $\xi$ values were derived using the \textit{Gaia}-ESO relation (Worley et al., in prep), calibrated for warm main-sequence stars:
\begin{equation}
\begin{split}
\xi =&1.10 + 6.04\,10^{-4}\cdot(T_{\rm{eff}}-5787) + 1.45\,10^{-7}\cdot(T_{\rm{eff}}-5787)^2 -\\ &-3.33\,10^{-1}\cdot(\log g-4.14) + 9.77\,10^{-2}\cdot(\log g-4.14)^2 +\\
&+6.94\,10^{-2}\cdot(\rm{[Fe/H]}+0.33) + 3.12\,10^{-2}\cdot(\rm{[Fe/H]}+0.33)^2
\end{split}
\end{equation}

, which is valid for stars with $T\rm{_{eff}}\geq 5200$\,K and $\log g\geq$3.5\,dex. 
In all the calibrated relations used to derive $T\rm{_{eff}}$, the bolometric correction, and $\xi$, the input metallicity was assumed to be solar, which was later confirmed by the chemical abundances analysis. The input values of the atmospheric parameters for all the stars are reported in Tables \ref{TabTs} (i.e. $T_{\rm{eff,phot}}$) and \ref{abb} (i.e. $\log g_{\rm{phot}}$ and $\xi_{\rm{phot}}$).\\

\begin{table*}[]
\caption{Atomic line data. The references for the $\log gf$ values are reported in Column 5. EWs and abundances for the Sun are reported in Columns 6 and 7. The abundances are in the $\log$(X) scale. The full table is available at the CDS. A portion is shown here for guidance regarding its form and content. }   
\label{tab_line1}      
\centering  
\setlength\tabcolsep{13pt}
\small
\begin{tabular}{lcccccr }    
\toprule  
Element & $\lambda$ & E.P. & $\log gf$ & Ref. & EW$_{\odot}$ & $\log$(X)$_{\odot}$\\
& \footnotesize{(\AA)} & \footnotesize{(eV)} & & & \footnotesize{(m\AA)}\\
\midrule
Na\,{\sc i } & 4982.814 & 2.104 & $-$0.916 & \cite{GESMCHF} & 75.34 & 6.223 \\
Na\,{\sc i } & 5682.633 & 2.102 & $-$0.706 & \cite{GESMCHF} & 106.92 & 6.259 \\
... & ... & ... & ... & ... & ... & ... \\
\midrule

\bottomrule
\end{tabular}
\end{table*}

\begin{table*}
\caption{Atmospheric parameters and chemical composition of Gaia benchmarks stars. We also report the value from the literature (i.e. $\xi$ values from \citealt{2015jofre} -J15-, while $T_{\rm{eff}}$, $\log g$, [Fe/H] and [Ti/H] are from \citealt{2018jofre} -J18-) and from the \textit{Gaia}-ESO iDR5 catalogue (GESiDR5). The numbers of the lines we used in the analysis are in square brackets. The uncertainties are calculated as the quadratic sum of the $\sigma_1$ and $\sigma_2$ contributions (see Sec.\ref{error} for details). }
\label{tabGaia}
\centering
\small
\setlength\tabcolsep{10pt}
\begin{tabular}{l|ccccc}
\toprule
 & Sun & $\alpha$Cen\,A & $\tau$Cet & $\beta$Hyi & 18Sco \\
\midrule
$T\rm{_{eff,J18}}$\,(K) & 5771$\pm$1 & 5792$\pm$16 & 5414$\pm$21 & 5873$\pm$45 & 5810$\pm$80\\ 
$\log g\rm{_{J18}}$\,(dex) &4.44$\pm$0.00 & 4.30$\pm$0.01& 4.49$\pm$0.01 & 3.98$\pm$0.02 & 4.44$\pm$0.03\\
\rowcolor{yellow}
$\rm{\xi_{J15}}$\,(km\,s$^{-1}$) & 1.06$\pm$0.18 & 1.20$\pm$0.07 & 0.89$\pm$0.28 & 1.26$\pm$0.05 & 1.07$\pm$0.20 \\

[Fe/H]$\rm{_{J18}}$ (dex) & 0.03$\pm$0.05 & 0.26$\pm$0.08 & $-$0.49$\pm$0.03 & $-$0.04$\pm$0.06 & 0.03$\pm$0.03\\

[Ti/H]$\rm{_{J18}}$ (dex) & 0.00$\pm$0.07 & 0.21$\pm$0.04 & $-$0.17$\pm$0.07 & $-$0.07$\pm$0.04 & 0.05$\pm$0.03\\
\hline
$T\rm{_{eff,GESiDR5}}$\,(K) & 5734$\pm$62 & 5813$\pm$60 & 5344$\pm$59 & 5828$\pm$116 & 5776$\pm$58\\ 
$\log g\rm{_{GESiDR5}}$\,(dex) &4.45$\pm$0.12 & 4.35$\pm$0.11& 4.56$\pm$0.11 & 3.90$\pm$0.23 & 4.42$\pm$0.11\\
\rowcolor{yellow}
$\rm{\xi_{GESiDR5}}$\,(km\,s$^{-1}$) & 0.90$\pm$0.18 & 1.06$\pm$0.08 & 0.61$\pm$0.30 & 1.25$\pm$0.07 & 0.99$\pm$0.13 \\

[Fe/H]$\rm{_{GESiDR5}}$\,(dex) & 0.06$\pm$0.11 & 0.27$\pm$0.11 & $-$0.50$\pm$0.13 & $-$0.06$\pm$0.10 & 0.08$\pm$0.07 \\

[Ti/H]$\rm{_{GESiDR5}}$\,(dex) & 4.95$\pm$0.03$^{\ast}$ & 0.26$\pm$0.01 & $-$0.23$\pm$0.04 & $-0.05\pm$0.03 & 0.03$\pm$0.02\\
\hline
$T(J-K)$\,(K) & 5777$\pm$1& 5845$\pm$61& 5401$\pm$51 & 5702$\pm$80 & 5295$\pm$134\\
$\log g\rm{_{phot}}$\,(dex) & 4.43$\pm$0.01 & 4.32$\pm$0.04 & 4.47$\pm$0.03 & 3.91$\pm$0.04 & 4.28$\pm$0.48 \\
$\rm{\xi_{phot}}$\,(km\,s$^{-1}$) & 1.00$\pm$0.00 & 1.08$\pm$0.03 & 0.79$\pm$0.02 & 1.13$\pm$0.03 & 0.79$\pm$0.15 \\
$T_{\rm{eff}}$(Fe\,{\sc i})\,(K)\,[N$\rm{_{lin}}$] & 5777$\pm$100 [59] & 5845$\pm$75 [55]  & 5401$\pm$50 [59] & 5800$\pm$100 [57] &5800$\pm$100 [59]  \\
$T_{\rm{eff}}$(Fe\,{\sc i}+Ti\,{\sc i})\,(K)\,[N$\rm{_{lin}}$] & 5790$\pm$50 [94] & 5830$\pm$75 [89] & 5401$\pm$75 [98] & 5870$\pm$100 [87] & 5875$\pm$100 [100]   \\
$\log g\rm{_{spec}}$\,(dex) & 4.47$\pm$0.05 & 4.45$\pm$0.10 & 4.38$\pm$0.10 & 3.95$\pm$0.10 &  4.55$\pm$0.10 \\
\rowcolor{yellow}
$\rm{\xi_{spec}}$\,(km\,s$^{-1}$) & 1.00$\pm$0.10 & 1.09$\pm$0.20 &  0.89$\pm$0.15 & 1.35$\pm$0.10 & 1.15$\pm$0.15 \\

[Fe/H]$\rm{_{I}}$ \,(dex) & 7.44$\pm$0.04$^{\ast}$ & 0.23$\pm$0.02 & $-$0.44$\pm$0.02 & $-$0.09$\pm$0.02 & 0.06$\pm$0.02 \\

[Fe/H]$\rm{_{II}}$ \,(dex) & 7.45$\pm$0.03$^{\ast}$ & 0.21$\pm$0.05 & $-$0.44$\pm$0.05 & $-$0.09$\pm$0.05 & 0.05$\pm$0.04 \\

[Ti/H]$\rm{_{I}}$ \,(dex) & 4.92$\pm$0.05$^{\ast}$ & 0.26$\pm$0.05 & $-$0.20$\pm$0.04 & $-$0.09$\pm$0.02 & 0.09$\pm$0.03 \\

[Ti/H]$\rm{_{II}}$ \,(dex) & 4.93$\pm$0.04$^{\ast}$ & 0.25$\pm$0.04 & $-$0.19$\pm$0.04 & $-$0.08$\pm$0.04 & 0.08$\pm$0.05 \\
\hline
\end{tabular}
\begin{tablenotes}
\small
\item $^{\ast}$ The solar values are in log(X) scale. 
\end{tablenotes}
\end{table*}

\section{New approach in elemental abundance analysis}\label{S4}

To derive spectroscopic atmospheric parameters and element  abundances of our target stars, we used the local thermodynamic equilibrium (LTE) line analysis and synthetic spectrum code \texttt{MOOG}\footnote{https://www.as.utexas.edu/~chris/moog.html} (version 2017, \citealt{sne73}; \citealt{2011sob}). Abundances of Fe\,{\sc i}, Fe\,{\sc ii}, Ti\,{\sc i}, Ti\,{\sc ii}, Na\,{\sc i}, Mg\,{\sc i}, Al\,{\sc i}, Si\,{\sc i}, Ca\,{\sc i}, Cr\,{\sc i,} and Ni\,{\sc i} were estimated using the EW method with the \textit{abfind} driver.

We used 1D model atmospheres that we linearly interpolated from MARCS grid  \citep{gusta08}, in the assumption that LTE and plane-parallel geometry is valid for dwarf stars. We chose these atmosphere models to be consistent with the analysis of the UVES spectra performed by the \textit{Gaia}-ESO consortium. The lines we used were taken from \cite{2017dorazi}, originally selected from the line list optimised for solar-type stars from \cite{2014melendez}. We cross-matched our original line list with the official \textit{Gaia}-ESO line list \citep{Heiter2019} to adopt the same atomic parameters, in particular the value for the oscillator strength ($\log gf$). The complete line list can be found in Table \ref{tab_line1}. We used the Barklem prescriptions for damping values (see \citealt{barklem2000} and references therein). 

The EWs for all the lines were measured using the software ARESv2 \citep{2015sousa}\footnote{http://www.astro.up.pt/~sousasag/ares/}. We discarded all the lines with uncertainties larger than 10\% and those lines with EWs$>$120\,m\AA\,because stronger lines cannot be fitted with a Gaussian profile. In some cases, especially for stars with relatively high rotational velocities, we added lines by measuring their EWs by hand using the task \textit{splot} in IRAF.\\ 

\begin{figure*}
\centering
\begin{minipage}[c]{19cm}
\centering
\includegraphics[scale=0.25]{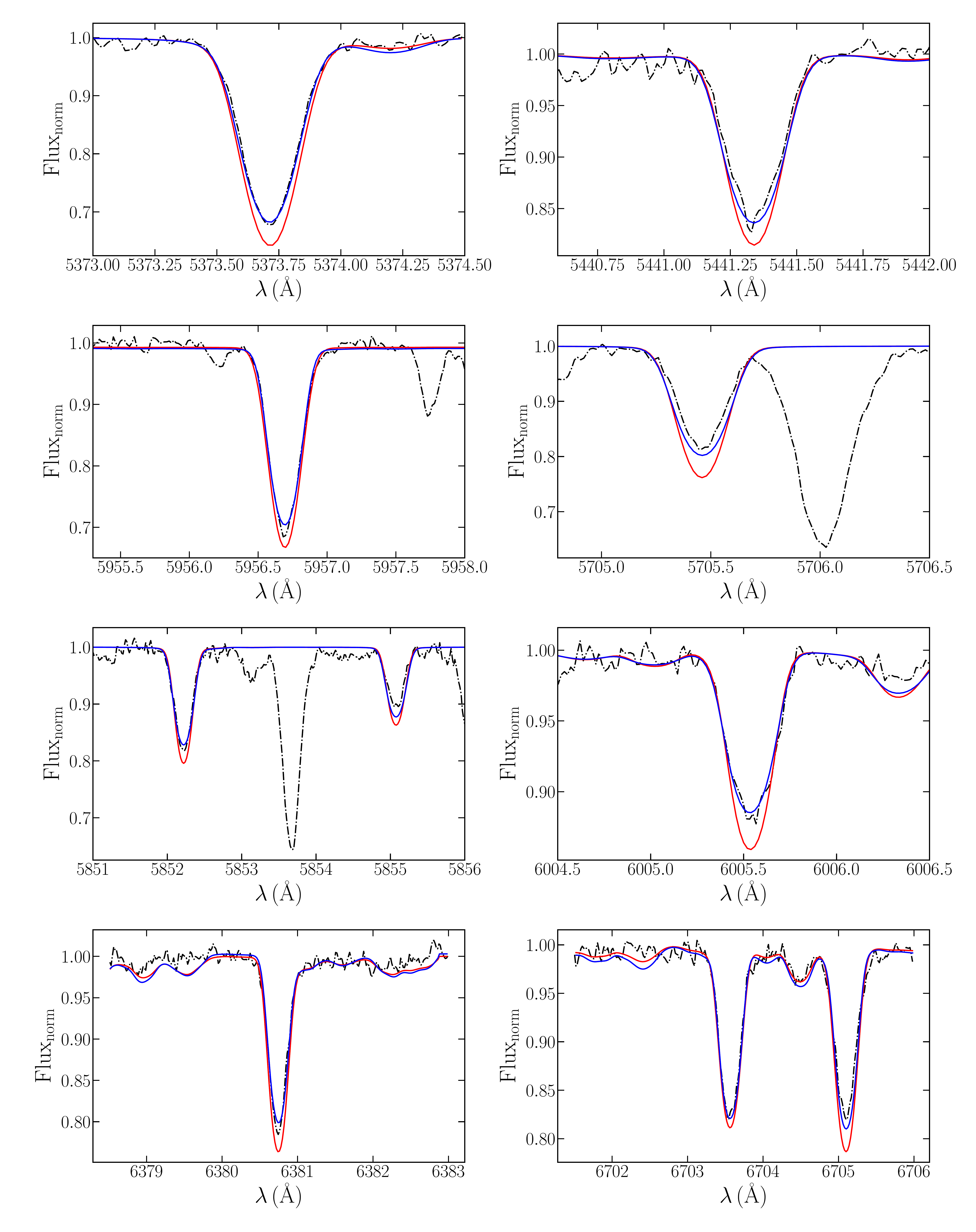}
\caption{Examples of the synthesis of several Fe\,{\sc i} lines of star 08365498$-$5308342 ($v\,{\rm sin}\, i = 9.88$~km\, s$^{-1}$, as the recommended value given in the iDR5) for which the observed profile is not fitted when we adopt the $\xi$ value obtained from the Fe lines ($\xi$=1.75\,km\,s$^{-1}$ ; red line) and with the values derived with the new method ($\xi$=0.85\,km\,s$^{-1}$ ; blue line).}
\label{synth}
\end{minipage}
\end{figure*}

\subsection{Standard analysis: iron lines}

First, we applied the \textit{\textup{standard}} analysis, which only uses neutral and ionised Fe lines to derive $T_{\rm{eff}}$ from the excitational equilibrium, $\log g$ by imposing the ionisation equilibrium, and $\xi$ by zeroing the trend with the REWs. We analysed the Sun and the Gaia FGK benchmark stars $\alpha$\,Cen\,A, $\tau$\,Cet, $\beta$\,Hyi, and 18\,Sco, whose UVES spectra were taken from \cite{blanco}. The atmospheric parameters for the Gaia benchmark stars are listed in Table \ref{tabGaia}. We selected only these four benchmark stars because their atmospheric parameters are similar to those of the stars in our sample. In addition, we analysed one star of the first\textup{} calibrator cluster IC\,2391 (age $\sim 50$\,Myr), 08365498$-$5308342. We obtain $T_{\rm{eff}}$=5350$\pm$100\,K, $\log g$=4.51$\pm$0.15\,dex,  $\xi$=+1.75$\pm$0.10\,km\,s$^{-1}$ , and [Fe/H]=$-0.07$. These values are similar to the iDR5 results, which are $T_{\rm{eff}}$=5381$\pm$55\,K, $\log g$=4.49$\pm$0.06\,dex, $\xi$=+1.77$\pm$0.01\,km\,s$^{-1}$ , and [Fe/H]=$-0.09$.

We investigated the nature of this quite high $\xi$ value in more detail by synthesising  all the Fe lines with the code \texttt{MOOG}. Fig.\ref{synth} shows that the synthetic profile with the anomalous value of $\xi$ (in red) does not reproduce the observed line profile (in black) in the weak or strong lines, which confirms our suspicion that the $\xi$ value is too high. This behaviour is confirmed in 90\% of the Fe lines of the line list. Instead, the synthetic profile with the atmospheric parameters, in particular $\xi$=0.85$\pm$0.10\,km\,s$^{-1}$, that we derived with the new method described in Sect. \ref{s4.2} (blue line) reproduces the observed profile better. \\ 

\begin{figure}
\centering
\includegraphics[width=0.47\textwidth]{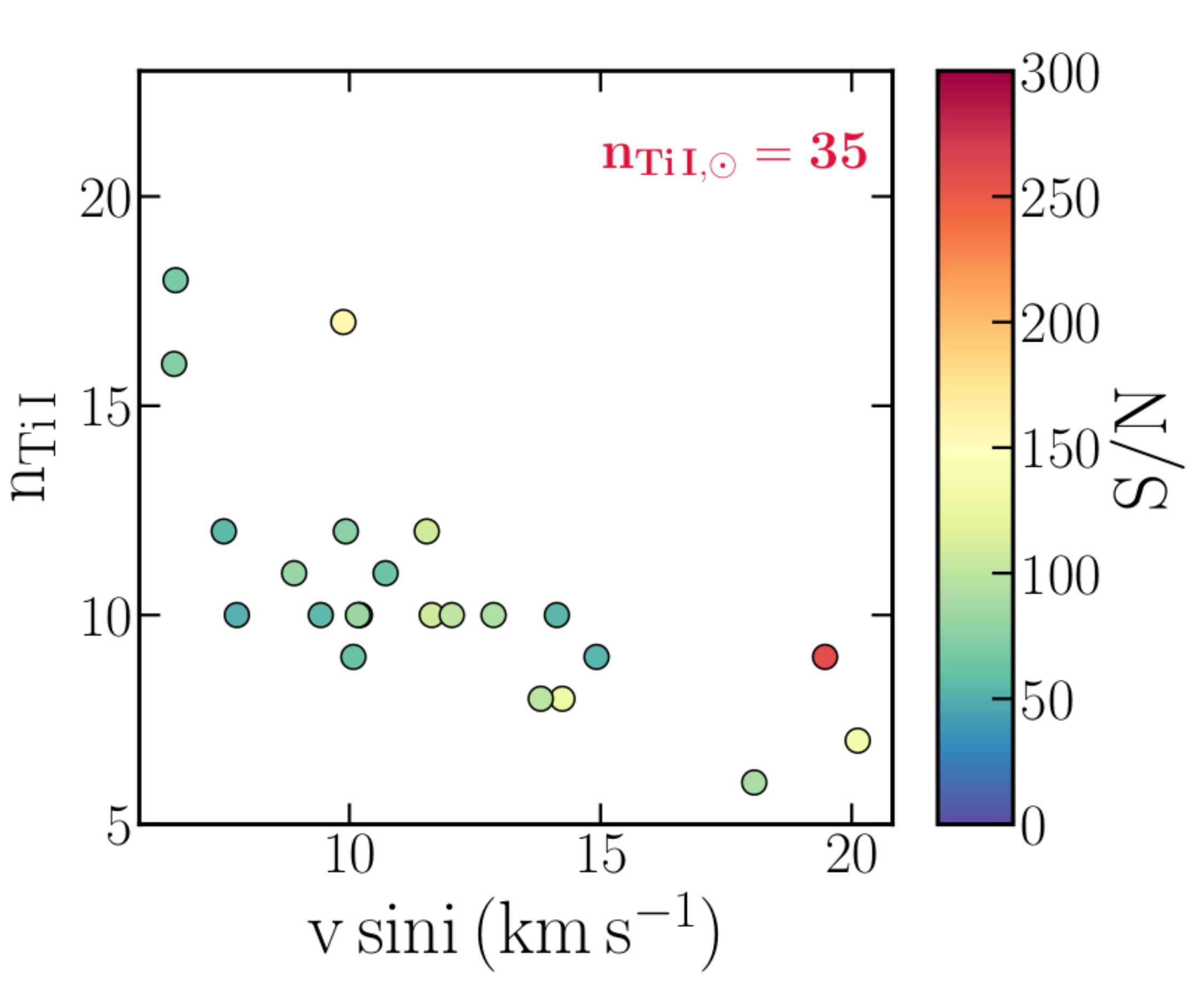}
\caption{Number of Ti\,{\sc i} lines measured for each star as a function of S/N, colour-coded according to $v\,{\rm sin}\, i < 20$~km\, s$^{-1}$. Stars with similar S/N but higher rotational velocities have fewer measurable lines. The number of Ti\,{\sc i} lines measured in the Sun is reported in the top right corner. }
\label{nline}
\end{figure}

\subsection{Titanium and iron lines}\label{s4.2}

We derived the atmospheric parameters using the second element with the largest number of spectral lines measurable in our stellar types, both of the neutral and the ionised species: titanium. On average, Ti lines form deeper in the photosphere than the Fe lines, at $\log\,\tau_{5000}\sim-1$, where $\log\,\tau_{5000}$ is the optical depth expressed in the logarithmic scale and calculated at the 5000\,$\AA$ reference wavelength. We therefore expect little influence from the chromosphere, that is, a lack of trends between abundances and line formation depth. Moreover, we have very precise laboratory measurements of the $\log gf$ values from \cite{2013lawler} for the Ti lines. Recently, \cite{tsa} have argued that especially for cool dwarf stars, the atmospheric parameter values derived with Ti lines are more reliable than those derived with Fe lines.
However, the authors used Ti lines only to impose the ionisation balance, in order to infer surface gravity estimates.
Here we expand upon this approach.
\\
The excitation potential (E.P.) range covered by the Ti lines used here is unfortunately too narrow to obtain a reliable estimate of $T_{\rm{eff}}$ (from 0 to 2.0 eV, while for Fe the range is 0-5.0 eV). All $T_{\rm{eff}}$ values we obtained with Ti lines alone are higher than the photometric estimates (200-300K higher), and the uncertainties are of about 200 K. Moreover, in some stars, especially those with higher rotational velocities, the number of measurable Ti lines is very low, in some cases there are even fewer than ten lines, as we show in Fig.\ref{nline}.\\
To obtain more lines and a wider coverage of the E.P. range, we derived $T_{\rm{eff}}$ values using Fe\,{\sc i} and Ti\,{\sc i} lines \textit{\textup{simultaneously}}. Tables \ref{tabGaia} and \ref{TabTs} show that the agreement between the three different estimates of $T\rm{_{eff}}$ is very good. Instead, for the other parameters we used Ti lines alone for the reasons presented above.\\

In summary, our new approach consists of the following three steps:
\begin{itemize}
\item deriving $T_{\rm{eff}}$ by zeroing the trend between E.P. and abundances of Ti\,{\sc i} and Fe\,{\sc i} lines \textit{\textup{simultaneously}}  (in particular, the slope of the trend is lower than the uncertainty on the slope, and the trend is not statistically meaningful);
\item deriving $\log g$ by imposing the ionisation equilibrium for Ti lines alone, that is, the difference between Ti\,{\sc i} and Ti\,{\sc ii} is of the order of the quadratic sum of the uncertainties calculated by \texttt{MOOG} divided by the square root of the number of lines of the two species;
\item deriving $\xi$ by zeroing the trend between the REW of Ti I lines alone and the abundances (as for the temperature, the slope of the trend is expected to be lower than the uncertainty on the slope).
\end{itemize}

\begin{figure}
\centering
\includegraphics[width=0.48\textwidth]{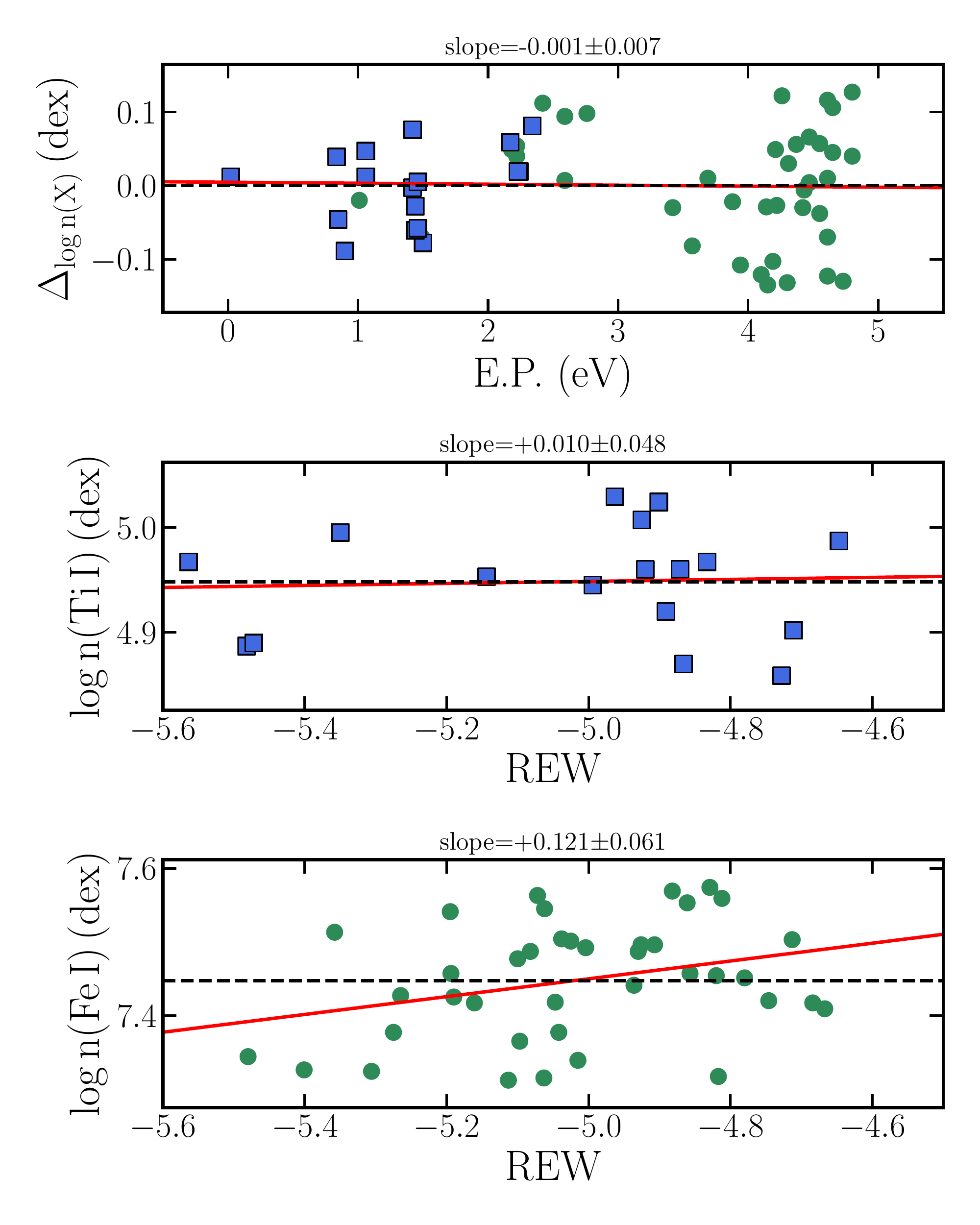}
\caption{Example of applying the new method for star 08365498$-$5308342 (IC\,2391, age 50\,Myr ). The green dots represent the Fe\,{\sc i} lines, and the blue squares represent the Ti\,{\sc i} lines; the red lines in both panels are the linear regression, and individual slope and uncertainties are reported for each panel. See text for further details. }
\label{example}
\end{figure}

When the star had fewer than ten measurable Ti\,{\sc i} lines, we kept the $\xi$ value fixed to the photometric estimate (7 of 23 stars). We considered 2 stars with discrepant $\xi$ values:  17452508$+$0551388 (IC\,4665, age 40\,Myr) and 08110139$-$04900089 (NGC\,2547, age 50\,Myr). We re-derived the atmopsheric parameters keeping the $\xi$ value fixed to the photometric guess. We find that for 17452508$+$0551388 the new parameters are equal to $T_{\rm{eff}}$=5321$\pm$150\,K,  $\log g$=4.35$\pm$0.10\,dex, and $\xi$=0.75$\pm$0.04\,km\,s$^{-1}$, while for 08110139$-$04900089, we find $T_{\rm{eff}}$=5325$\pm$100\,K, $\log g$=4.47$\pm$0.10\,dex, and $\xi$=0.80$\pm$0.04\,km\,s$^{-1}$. The difference between the newly derived $T_{\rm{eff}}$ and those we derived with our method is lower than 50\,K and lower than the typical error (75-100\,K) we obtain for the $T_{\rm{eff}}$ parameter. We therefore conclude that the two $T_{\rm{eff}}$ measurements are consistent and that the effect of these differences on the final metallicity values is weak.  For the Sun, we measure 11 Ti\,{\sc ii} lines, and for our sample stars we measure from 4 to 8 lines, depending on the quality of the spectra. Fig.\ref{example} shows an example of the trends with the final parameters for star 08365498$-$5308342. In the y-axis of the top panel, the difference between individual lines and the mean values per atomic species is reported and was calculated as  $\rm{\Delta_{log\,n(X)}=log\,n(X)_i-{log\,n(X)_{mean}}}$.  The trend between Fe\,{\sc i} line abundances and REWs clearly suggests that the $\xi$ value needs to be further increased. \\
All the final abundances were calculated differentially with respect to the Sun.
The final model atmospheres and abundances are reported in Table \ref{abb}. 
We also calculated the abundances for other different $\alpha$- and proton-capture and iron-peak elements, in particular, Na\,{\sc i}, Mg\,{\sc i}, Al\,{\sc i}, Si\,{\sc i}, Ca\,{\sc i}, Cr\,{\sc i,} and Ni\,{\sc i}. The respective abundance ratios [X/Fe] were calculated as [X/Fe]=[X/H]$_{\star}-$[Fe/H]$_{\star}$ ( in particular, [Ti/Fe]$_{\rm{II}}$ = [Ti/H]$_{\rm{II}}-$[Fe/H]$_{\rm{II}}$). The final abundance ratios are reported in Table \ref{ratios}.\\

\begin{table}[]
\caption{Solar abundances derived here and in \cite{aspl} (A09), and meteoritic abundances from \cite{lodd} (L09). We also report the values derived by \textit{Gaia}-ESO in iDR5 (GESiDR5).}
\centering
\begin{tabular}{lcccc}
\toprule
Species & This work & A09 & L09 & GESiDR5\\
\midrule
Na & 6.24$\pm$0.04 & 6.24$\pm$0.04 & 6.27$\pm$0.02 & 6.17$\pm$0.05 \\
Mg & 7.63$\pm$0.02 & 7.60$\pm$0.04 & 7.53$\pm$0.01 & 7.51$\pm$0.07 \\
Al & 6.43$\pm$0.03 & 6.45$\pm$0.03 & 6.43$\pm$0.01 & 6.34$\pm$0.04 \\
Si & 7.47$\pm$0.01 & 7.51$\pm$0.03 & 7.53$\pm$0.01 & 7.48$\pm$0.06 \\
Ca & 6.27$\pm$0.04 & 6.34$\pm$0.04 & 6.29$\pm$0.02 & 6.31$\pm$0.12 \\
Ti & 4.93$\pm$0.01 & 4.95$\pm$0.05 & 4.91$\pm$0.03 & 4.95$\pm$0.06 \\
Cr & 5.57$\pm$0.03 & 5.64$\pm$0.04 & 5.64$\pm$0.01 & 5.61$\pm$0.09 \\
Fe & 7.45$\pm$0.01 & 7.50$\pm$0.04 & 7.45$\pm$0.01 & 7.49$\pm$0.03 \\
Ni & 6.19$\pm$0.04 & 6.22$\pm$0.04 & 6.20$\pm$0.01 & 6.23$\pm$0.07 \\
\hline
\end{tabular}
\label{solar}
\end{table}

\subsection{Solar abundance scale and the Gaia benchmark stars}

We applied our new method to the Gaia benchmark stars to determine its validity, and to the solar spectrum to derive our solar abundance scale. In Table \ref{tabGaia} we report the values of the atmospheric parameters and abundances of the Gaia benchmark stars. The results we obtain with our new method are in excellent agreement with the results by \cite{2015jofre,2018jofre} and with \textit{Gaia}-ESO, in particular for $\xi$ (rows highlighted in yellow). We obtain very similar results, which confirms our hypothesis that the standard analysis produces good results for older stars ($\gtrsim 600\,$Myr).  

In Table \ref{solar} we report the solar abundance scale, obtained with the atmospheric values reported in Table \ref{tabGaia}. The mean value between Fe\,{\sc i} and Fe\,{\sc ii} for the final Fe abundance is reported, also for Ti. The uncertainties are the quadratic sum of the $\sigma_1$ and $\sigma_2$ contributions (calculated as the scatter measured by \texttt{MOOG} divided by the square root of the number of lines and as the sensitivity of the abundances to uncertainties in the atmospheric parameters, respectively). Our solar abundances generally agree well with the results by \cite{aspl}, with the meteoritic results by \cite{lodd}, and also with the results of \textit{Gaia}-ESO iDR5. Based on the study of \cite{2011bergemann} on the NLTE effects on Ti, we expect that NLTE corrections for our lines that form at $\log(\tau_{5000})\sim-1$ and at solar metallicities are negligible.  \\

\subsection{Uncertainty estimates}\label{error}

The uncertainties of the atmospheric parameters reported in Tables \ref{TabTs} and \ref{abb} were estimated as follows. $\sigma_{{T{_{\rm eff}}}}$ was calculated by varying the temperature until the slope E.P. versus abundance was larger than its uncertainty and the trend became statistically meaningful. $\sigma_{\log g}$ was estimated by varying $\log g$ until $\Delta$(Ti\,{\sc ii}$-$Ti\,{\sc i}) was larger than the quadratic sum of the uncertainties. Finally, $\sigma_\xi$ was calculated by varying $\xi$ until the slope of the trend was larger than the uncertainty on the slope and the trend became statistically meaningful.  We individually evaluated the uncertainties for each star; in general, they are about 75-100\,K, 0.10\,dex, and 0.10-0.15\,km\,s$^{-1}$ for $T\rm{_{eff}}$, $\log g,$ and $\xi$, respectively.

The uncertainties on the abundances, $\sigma_1$ and $\sigma_2$ reported in Table \ref{abb}, take the internal uncertainties and the contribution of the atmospheric parameters into account, respectively.
The first source of uncertainty, $\sigma_1$, can be represented by the standard deviation from the mean abundance considering all the lines divided by the square root of the number of lines.
The $\sigma_2$ values instead represent the sensitivity of [X/H] to the uncertainties in the atmospheric parameters, and this sensitivity is calculated as
$$
\sigma_2=\sqrt{\left( \sigma_{T\rm{_{eff}}}  \frac{\partial \rm{[X/H]}}{\partial T\rm{_{eff}}}\right) ^2 + \left( \sigma_{\rm{\log\,g}} \frac{\partial \rm{[X/H]}}{\partial \rm{\log g}}\right) ^2 +\left( \sigma_{\xi} \frac{\partial \rm{[X/H]}}{\partial \rm{\xi}} \right) ^2}. 
$$

As reported in Table A.3, for the uncertainties of the abundance ratios [X/Fe], the $\sigma_1$ values were calculated by quadratically adding the $\sigma_1$ value of [Fe/H] and that of [X/H]. The $\sigma_2$ values were instead calculated in the same way as for [X/H].\\


\begin{figure*}[h!]
\centering
\begin{minipage}[c]{19cm}
\centering
\includegraphics[width=\textwidth]{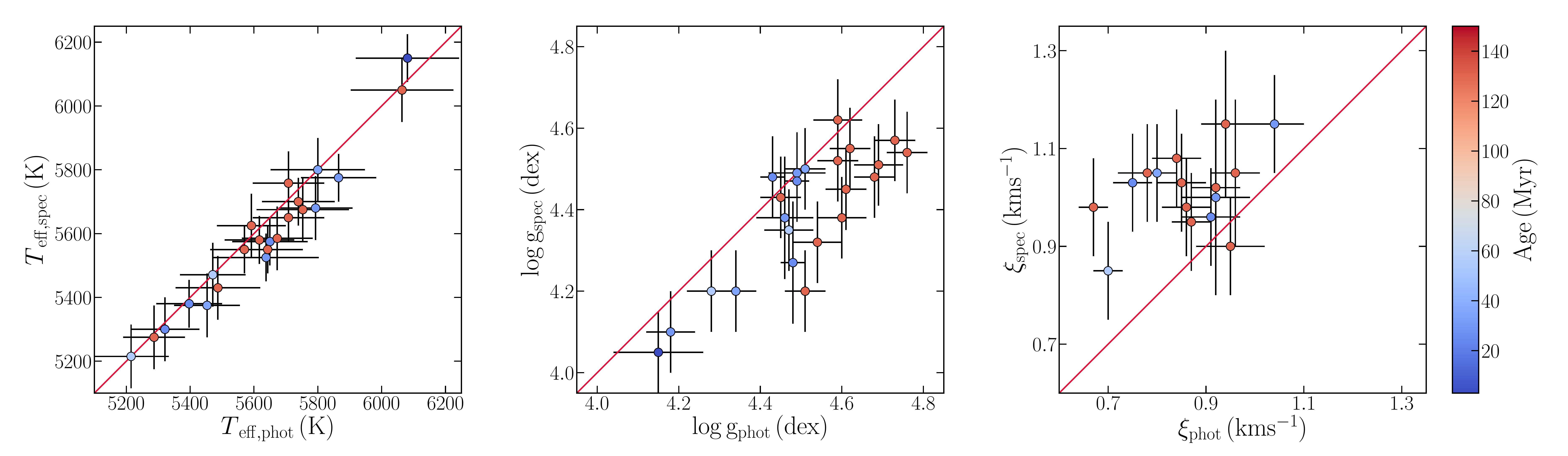}
\caption{Comparison between the photometric and spectroscopic estimates of the atmospheric parameters. The red line represents the 1:1 relation. The points are colour-coded according to ages. The red points refer to NGC\,2516, and the blue points to the younger clusters.}
\label{conf_fot_spec}
\end{minipage}
\end{figure*}

\section{Results and discussion}\label{S5}

\subsection{Stellar atmospheric parameters}

\begin{figure*}
\centering
\begin{minipage}[c]{19cm}
\includegraphics[width=\textwidth]{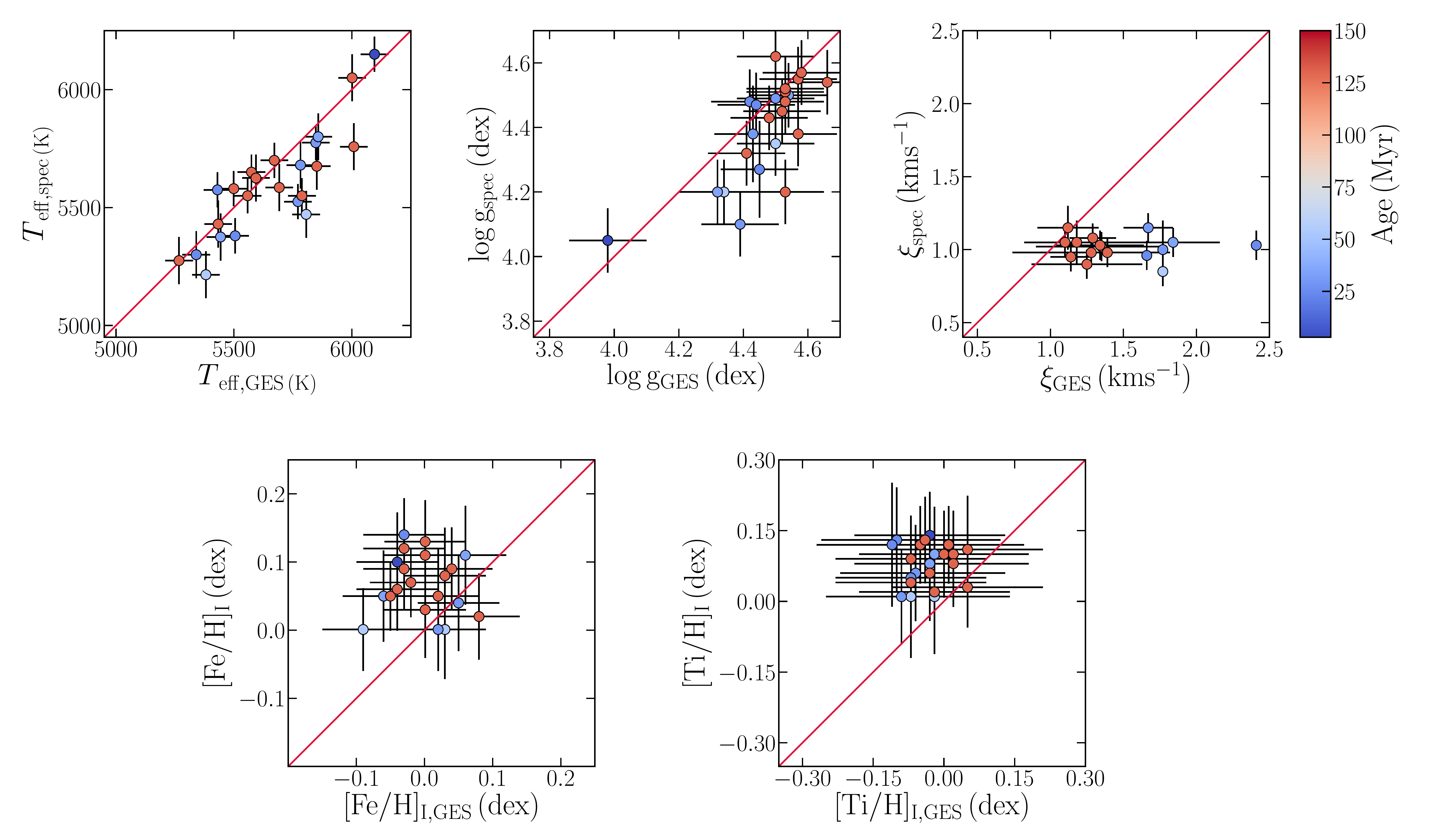}
\caption{ Comparison between the values of $T\rm{_{eff,spec}}$, $\log g_{\rm{spec}}$, $\xi_{\rm{spec}}$ , and [X/H] derived here  with the values from \textit{Gaia}-ESO iDR5. The red line represents the 1:1 relation. We exclude stars from the plot whose $\xi$ value is fixed to the photometric estimates. See text for further details. }
\label{conf_ges_spina}
\end{minipage}
\end{figure*}

The atmospheric parameters we inferred with our method agree well with the photometric estimates (Fig.\ref{conf_fot_spec}, where the data are colour-coded according to age). The temperatures agree very well, with a mean difference of $-$37$\pm$49\,K. We note a general offset of the $\log g$ values, with a difference of $-$0.11$\pm$0.09\,dex. However, we also note that the ionisation equilibrium is valid for both Ti and Fe (Table \ref{abb}). For the $\xi$ values, we find instead that the spectroscopic estimates are slightly larger than the photometric ones, but still comparable, with a difference of 0.15$\pm$0.10\,km\,s$^{-1}$. We excluded all stars from the comparison plots whose $\xi$ value was fixed to the photometric estimate. However, this offset between spectroscopic and photometric determinations of $\xi$ is known in the literature and has also been observed in old stars. We note that using the values derived by \textit{Gaia}-ESO ($\sim2.0$\,km\,s$^{-1}$), the difference with the photometric estimates is even larger. 

We compared our results of the atmospheric parameters and the Fe and Ti abundances with those given in the iDR5 catalogue. The comparison plots are shown in Fig.~\ref{conf_ges_spina} (as in Fig. \ref{conf_fot_spec}, the data are colour-coded by age): our measurements of $T\rm{_{eff}}$ and $\log g$ are in fair agreement with \textit{Gaia}-ESO. For $T_{\rm{eff}}$ we obtain  mean differences between our values and \textit{Gaia}-ESO results of $-$66$\pm$122\,K. Instead, for $\log g$ we found $\Delta \log g=-0.07\pm$0.11\,dex. We conclude that our results are reliable and agree with those of \textit{Gaia}-ESO. However, the largest differences are seen for the $\xi$ parameter. Our results are lower than those of iDR5. We find a mean difference of $-$0.46$\pm$0.36\,km\,s$^{-1}$. A small mean difference like this can be explained by the fact that for the stars in NGC\,2516 ($\tau\sim130$\,Myr), we obtain values of $\xi$ that agree with iDR5. 

The third panel in the top row in Fig.\ref{conf_ges_spina} shows a net separation between younger OCs and older OCs (in this case, only NGC\,2516, represented by the red points). The most dramatic effect of over-estimating the $\xi$ value is seen in the youngest clusters. We calculated the mean differences for the clusters with ages younger than 100\,Myr and for NGC\,2516 with ages older than 100\,Myr. While the oldest cluster has a mean difference of $-$0.23$\pm$0.13\,km\,s$^{-1}$, meaning that our results are comparable to the iDR5 results, for the youngest clusters we have a mean difference of $-$0.85$\pm$0.27\,km\,s$^{-1}$, which is significant at the 3$\sigma$ level. These results confirm our hypothesis that the standard analysis might produce over-estimated values of the $\xi$ parameter for young stars, which leads to an under-estimation of the element abundances, as is shown in the bottom panels of Fig.\ref{conf_ges_spina}.

It is noteworthy that for the NGC 2516 stars we derive a slightly higher metallicity than was published by \textit{Gaia}-ESO, although the $\xi$ values are quite similar.  In particular, even if the difference in $\xi$ is smaller than the difference with the younger stars, the range covered by the difference in [Fe/H] is the same as for the younger stars. This can be explained as the result of small differences in  the other photospheric parameters (e.g. 100\,K in temperature produces 0.07\,dex in metallicity), to the use of different criteria in zeroing the trends in order to derive the photospheric parameters, to the use of different method, such as the spectral synthesis, and to differences in the EW measurements. 

\begin{figure}[h]
\centering
\includegraphics[width=0.45\textwidth]{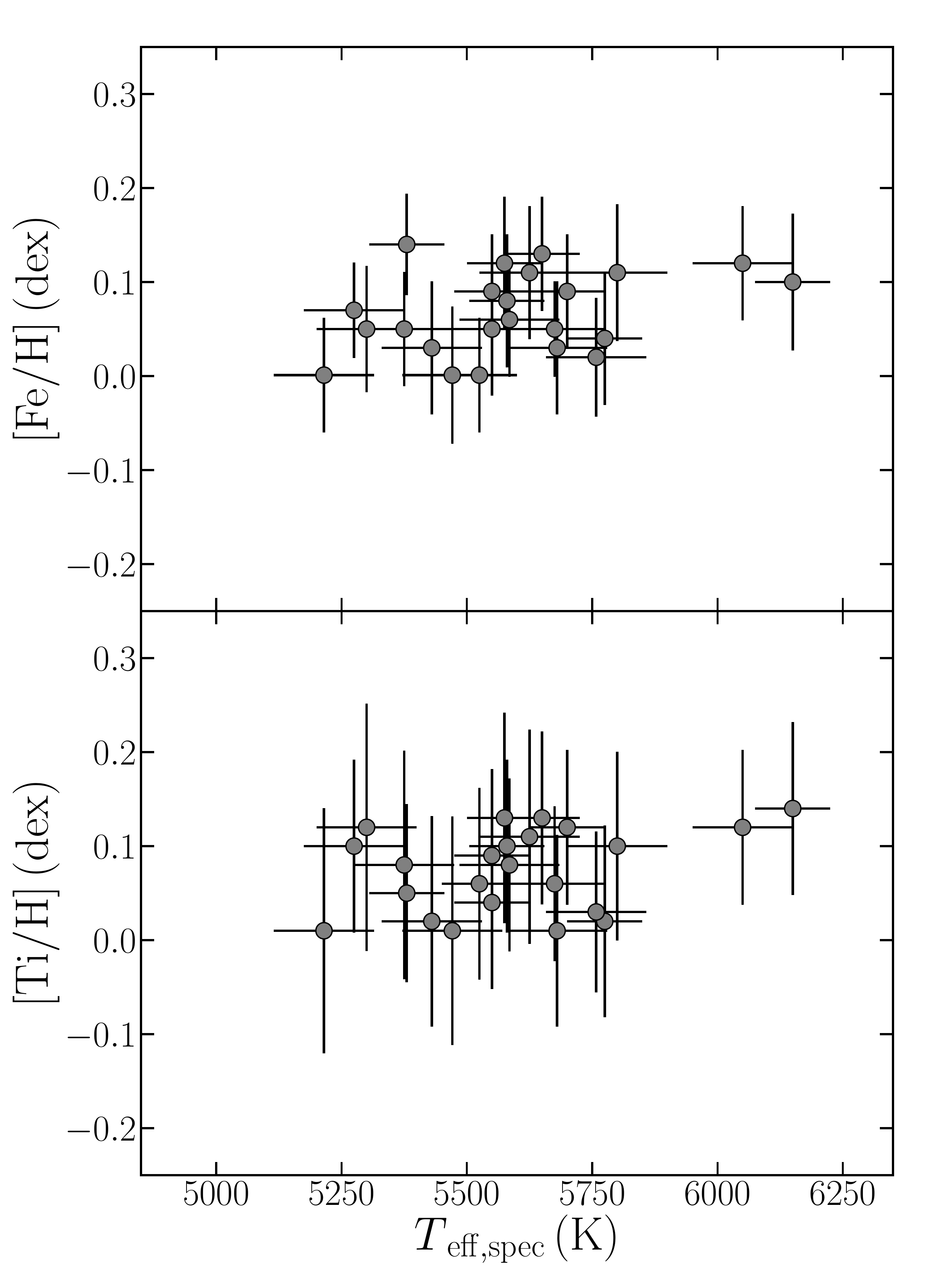}
\caption{[Fe/H] (top panel) and [Ti/H] (bottom panel) as a function of $T\rm{_{eff}}$ derived using Fe and Ti lines simultaneously. For Fe, the Pearson correlation coefficient is $r$=0.35, with $p$=.10, which is $\text{not}$ significant at $p<.05$. For Ti, the Pearson correlation coefficient is $r$=0.32, with $p=.14$, which is $\text{not}$ significant at $p<.05$.}
\label{abb_teff}
\end{figure}

\subsection{Element abundances}

Final parameters and abundance ratios are reported in Tables \ref{abb} and \ref{ratios}. The mean values for each cluster are reported in Table \ref{mean_Ab}, where the uncertainties represent the uncertainty on the mean. In some cases it was not possible to derive the abundances of some element, such as Al\,{\sc i}, because the lines were too weak to be measured or because of blending with nearby lines.

\begin{figure}
\centering
\includegraphics[width=0.5\textwidth]{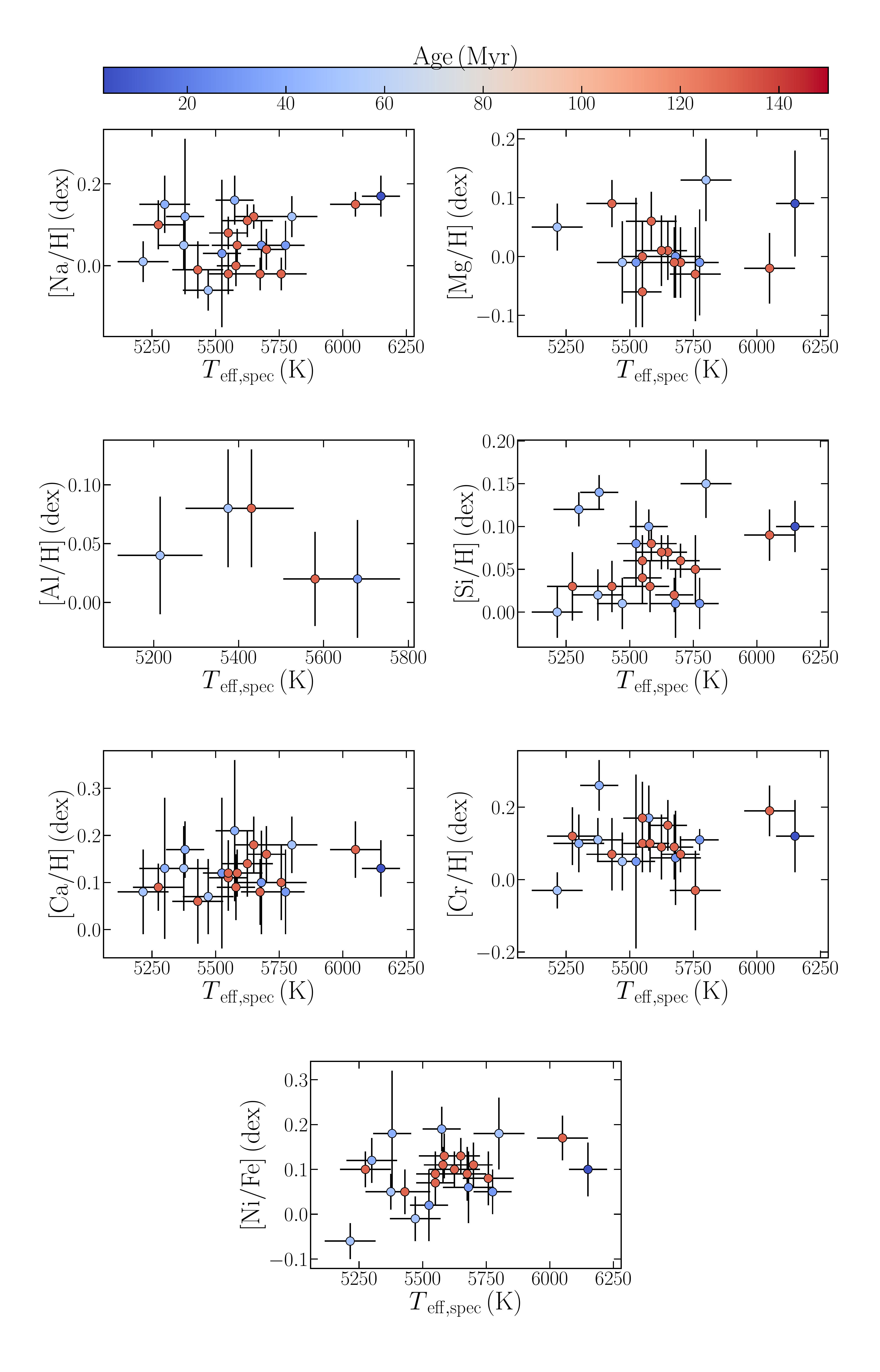}
\caption{[X/H] as function of $T\rm{_{eff}}$, derived with the new method and colour-coded according to age. }
\label{rapp_teff}
\end{figure}

\begin{figure}
\centering
\includegraphics[width=0.5\textwidth]{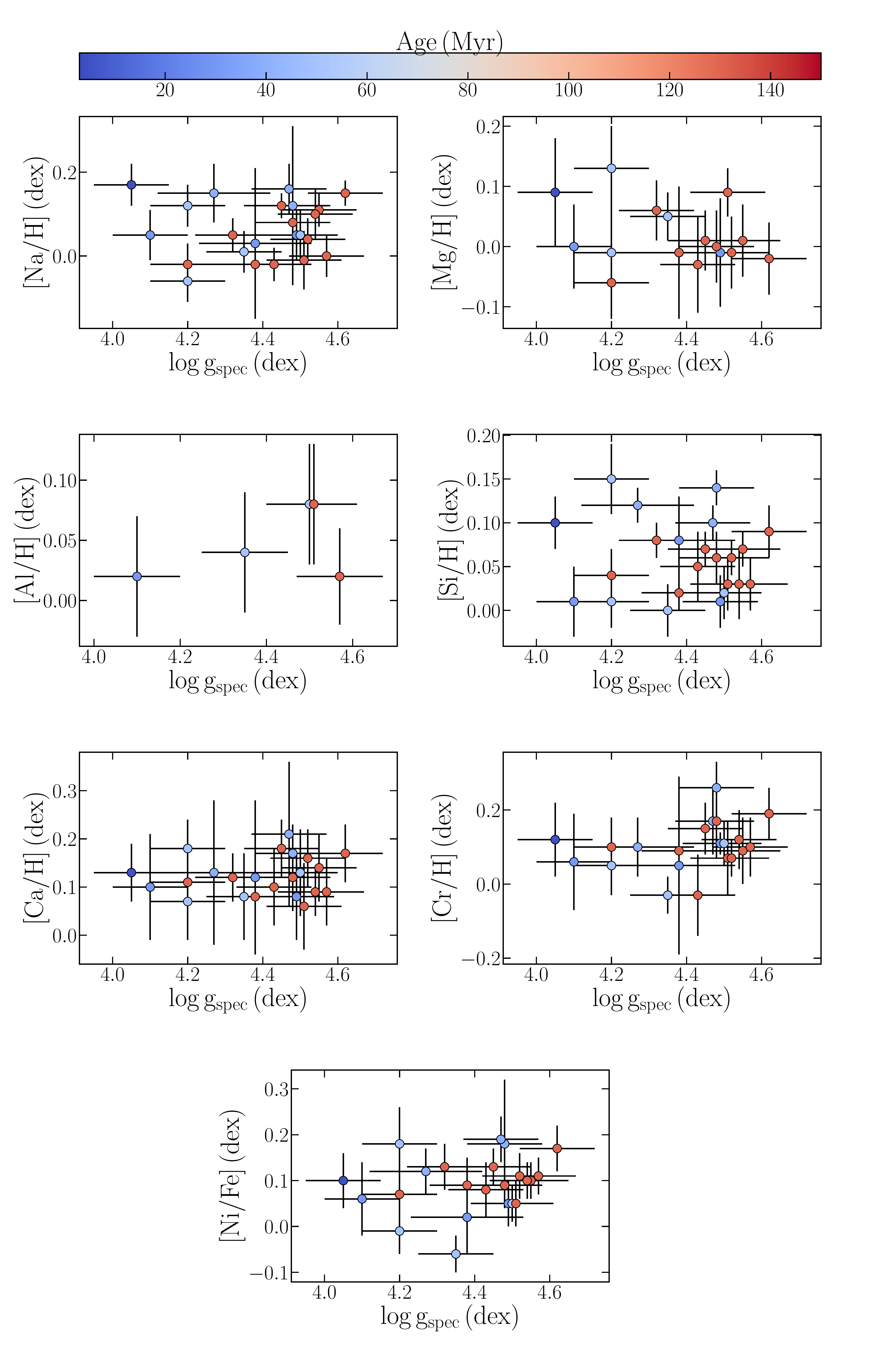}
\caption{[X/H] as function of $\log\,g$, derived using only Ti lines and colour-coded according to age. }
\label{rapp_logg}
\end{figure}

\begin{figure}
\centering
\includegraphics[width=0.4\textwidth]{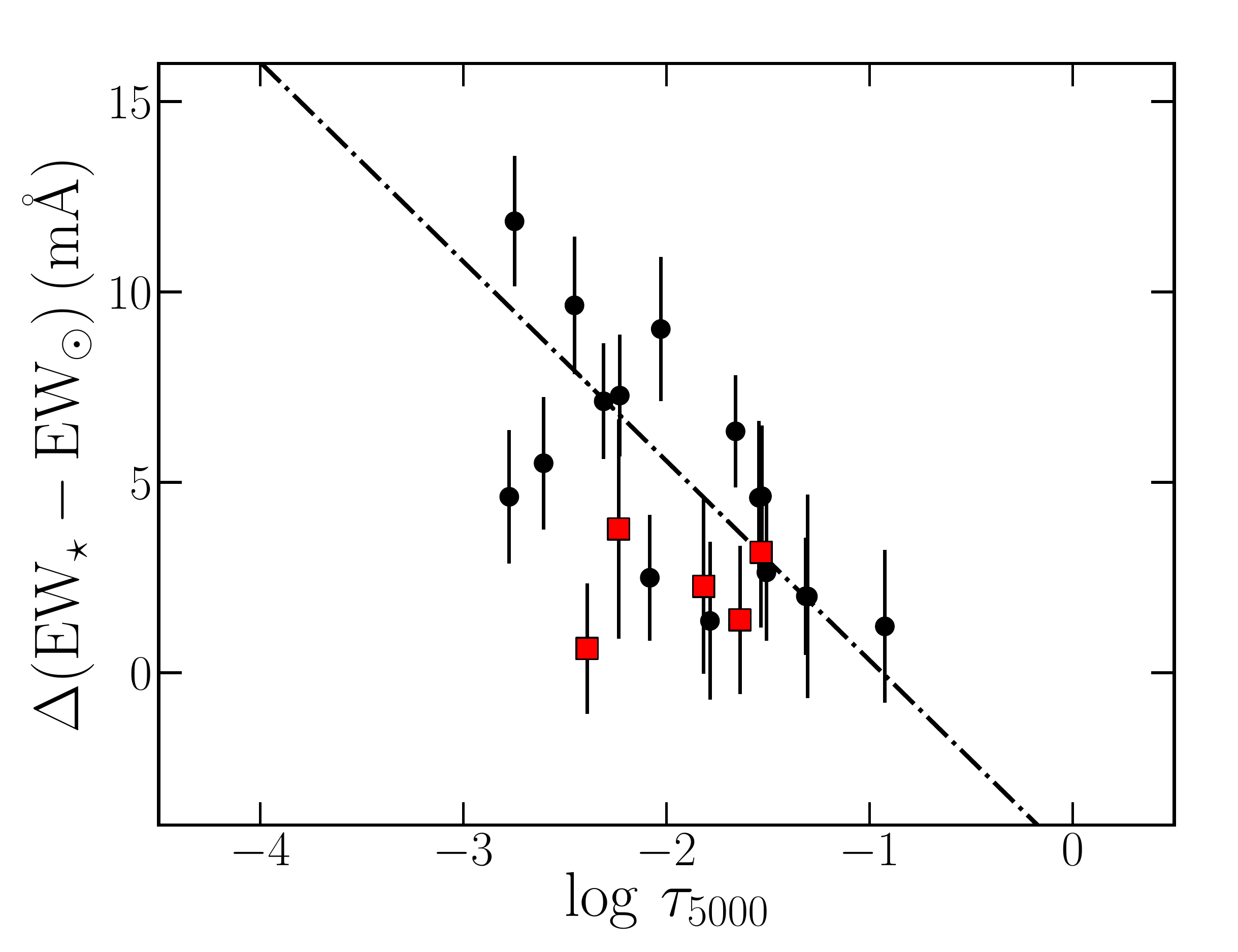}
\caption{Difference between the EWs of the Fe (black dots) and Ti (red dots) lines measured in the solar-analogue star 10442256$-$6415301 (30\,Myr) and the Sun as a function of the optical depth of line formation $\log\,\tau_{5000}$. The dot-dashed line is the trend for Fe lines.}  
\label{ew_tau}
\end{figure}

\begin{figure}
\centering
\includegraphics[width=0.4\textwidth]{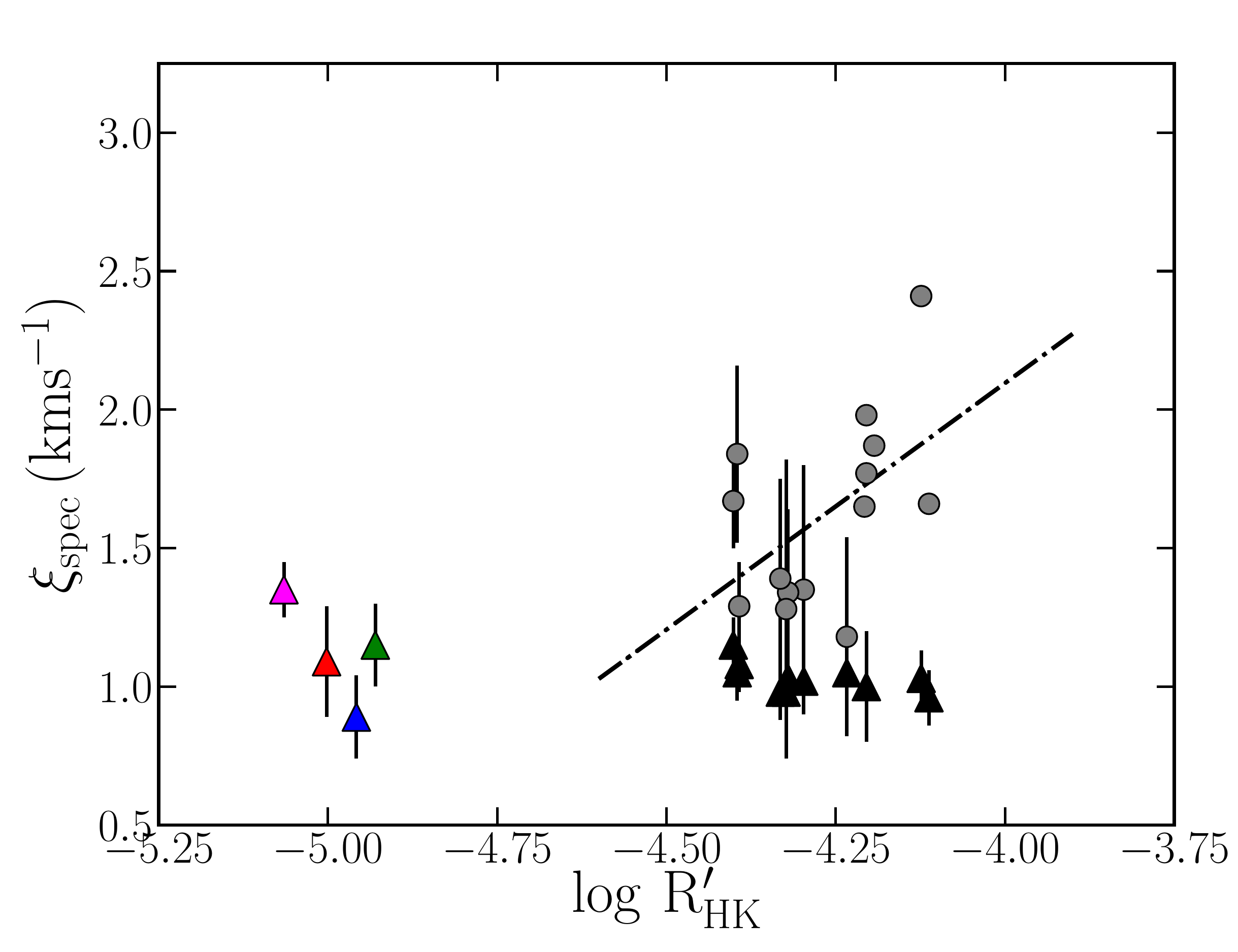}
\caption{Activity index $\log\,\rm{R^{\prime}_{HK}}$ as a function of $\xi$ values: the GESiDR5 results are represented with grey circles, and the triangles represent the values we find with our new method. The black triangles are the stars in the sample, and the coloured triangles represent the Gaia benchmark stars: red for $\alpha\,\rm{Cen\,A}$, green for 18\,Sco, blue for $\tau\,$Cet, and magenta for $\beta\,$Hyi. The dot-dashed line is the trend observed for the GES values.} 
\label{activity}
\end{figure}

\begin{figure*}[h!]
\centering
\includegraphics[scale=0.28]{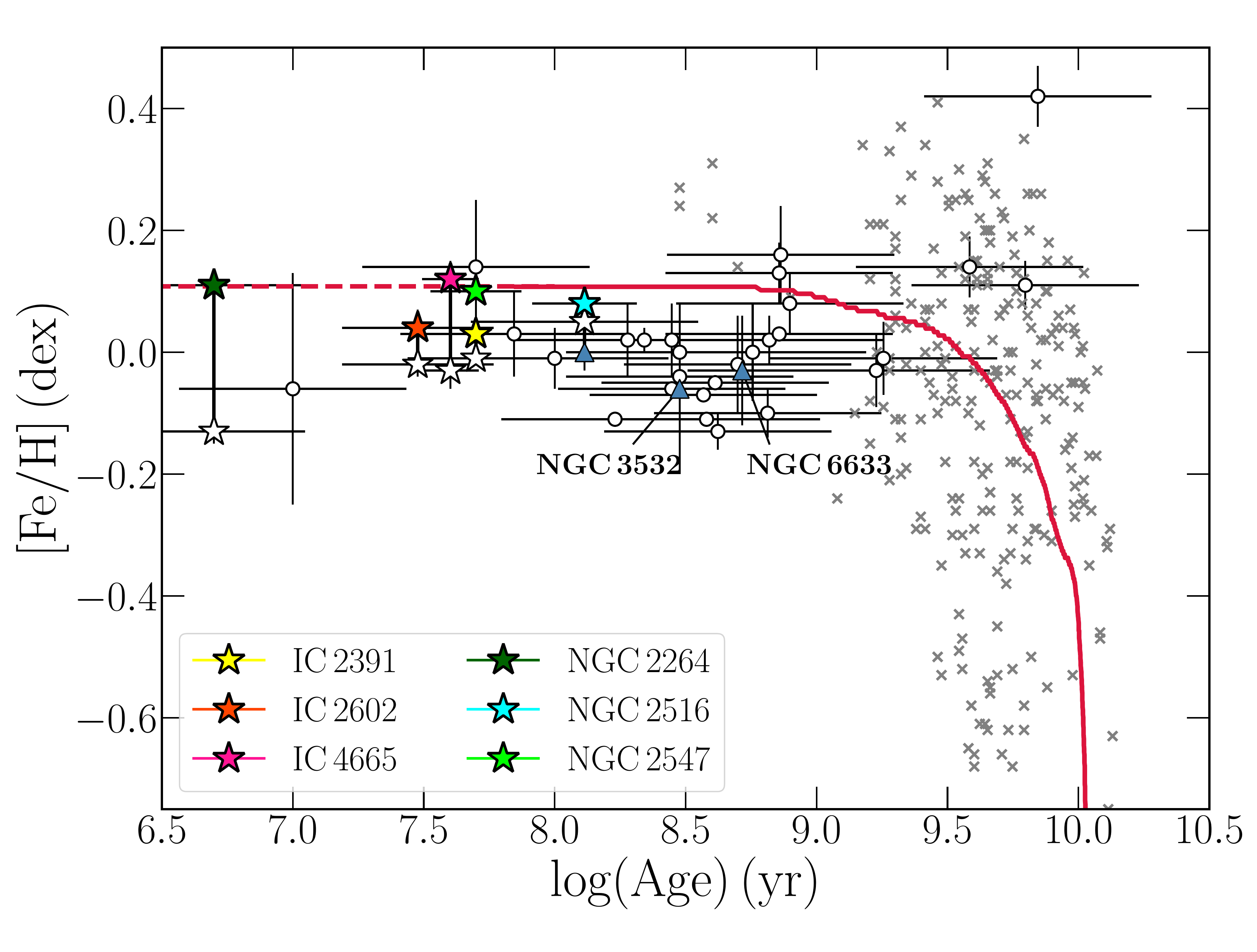}
\caption{Age-metallicity distribution for the YOCs we analysed here, for the sample of \cite{neto} (empty circles), field stars taken from \cite{ben} (grey crosses), and \textit{Gaia}-ESO clusters analysed in \cite{magrini} (filled triangles). The red line represents the model by \cite{minchev} for 7.5$< R\rm{_{Gal}} < 9$\,kpc. See text for further details. The empty stars represent the clusters we analysed whose metallicity was derived with the standard analysis \citep{neto}. } 
\label{age_met}
\end{figure*}

To evaluate the validity of our method, we determined the correlation between the derived abundances with T$\rm{_{eff}}$, as in Fig.\ref{abb_teff} for Fe and Ti and Fig.\ref{rapp_teff} for the different [X/H] ratios. We also determined the trends between [X/H] and $\log g$ (Fig.\ref{rapp_logg}). In these figures, the data are colour-coded according to the age, as for the previous plots. The $\alpha$- and proton-capture and the iron-peak elements elements overall show solar abundances, as expected for these types of objects. We do not find any statistically meaningful trend, therefore our results are expected to be reliable.

Originally, our sample included a few stars with $T_{\rm{eff}}\lesssim 5200$\,K. For these stars we find discrepancies (differences larger than 0.8\,dex) between abundances derived from Fe\,{\sc i} and Fe\,{\sc ii}, as well as Ti\,{\sc i} and Ti\,{\sc ii}. These can be explained with the so-called \textit{\textup{overionisation effect}}. It has been confirmed by different authors that stars with $T\rm{_{eff}} \lesssim 5200$\,K and young ages ($\tau \lesssim 100$\,Myr) show systematically larger Fe\,{\sc ii} abundances with respect to Fe\,{\sc i} in clusters and in field stars \citep{king,chen,schu,ben}. These differences are stronger as $T\rm{_{eff}}$ decreases and dramatically affect the derived atmospheric parameters values, in particular $\log g$. This is also valid for the Ti lines (see Fig.4 in \citealt{dor}). According to the results reported by \citealt{dor}, the over-ionisation effect reaches values up to 0.6\,dex for stars with $T\rm{_{eff}}$ lower than 5000\,K, which decrease with increasing $T\rm{_{eff}}$. Over-ionisation and/or over-excitation effects drove our choice to restrict the analysis to star with $T\rm{_{eff}} \gtrsim 5200$\,K.\\
\noindent
We compared our final mean results with literature values. For IC\,2391, IC\,2602, IC\,4665, and NGC\,2516, our measurements in general agree fairly well with different studies (\citealt{desilva}, \citealt{dor}, \citealt{shen}, and \citealt{ten}, respectively). For NGC\,2264, \cite{king} derived abundances for three stars and obtained a mean metallicity of $-$0.15$\pm$0.09. \cite{king} studied two stars with $T\rm{_{eff}}<5000$\,K and one star that was similar to the Sun. The authors derived $T\rm{_{eff}}$ and $\xi$ with the standard spectroscopic analysis, but they fixed $\log$\,g to the value estimated based on the isochrones. Their Table 1 lists $\xi$ values of $\sim$2.0\, km$s^{-1}$, which causes the Fe abundances to become sub-solar.  

\subsection{Effect of stellar activity}

To analyse the effect of stellar activity on Fe lines, we studied the dependence of the Fe line EWs on optical depth $\log\tau_{5000}$, taken from \cite{2015gurtovenko}. In Fig.\ref{ew_tau}, we calculate the difference of the EWs of the Fe and Ti lines between a solar analogue (star 10442256$-$6415301 belonging to IC2602, with an age of 30 Myr) and the Sun. The solar analogue has $T_{\rm{eff}}=5775\pm75\,$K and $\log g=4.49\pm0.10$\,dex, and we assumed that the metallicity is the same as the Sun. Lines forming in the upper layers of the atmosphere ($\log\tau_{5000}<-2.5$) in the young star have larger EWs than those in the Sun, with differences up to 5-10\,m$\AA$. The linear trend  for the Fe lines has a Pearson correlation coefficient $r=-0.85$ and is significant at $p<.01$. Lines forming in the external layers are much stronger in the young star than in the Sun; this means that this effect can influence the derivation of the $\xi$ parameter when it is derived based on the Fe lines. 

We also analysed the dependence of the $\xi$ parameter on the chromospheric activity index $\log\,\rm{R^{\prime}_{HK}}$. Because we cannot directly calculate $\log\,\rm{R^{\prime}_{HK}}$  from our spectra (because the spectral coverage does not include the Ca\,{\sc ii} H and K lines), we used the conversion relation found in \cite{2008mamajek}, which takes the $\log\,(L_{\rm{X}}/L_{\rm{bol}})$ activity index into account. We find in the literature values for 14 of the 23 stars we analysed. Fig.\ref{activity} shows that while our measurement  does not display a significant trend with activity, the $\xi$ values measured by \textit{Gaia}-ESO instead increase at increasing $\log\,\rm{R^{\prime}_{HK}}$, that is, at an increasing level of activity.  In this figure, we also report the values for the Gaia benchmark stars we analysed. The benchmarks are quiet stars, with  $\log\,\rm{R^{\prime}_{HK}}< -4.8$  , and we obtain the same value as \cite{2015jofre} with the new approach, as reported in Table \ref{tabGaia}. We also note that the $\xi$ value we obtain for $\beta$\,Hyi is slightly higher than the values obtained for the other benchmark stars, which is mainly due to the slightly advanced evolutionary stage.

\section{Concluding remarks}\label{S6}

We proposed a new approach to deriving the stellar atmospheric parameters. We analysed a sample of 23 dwarf stars observed in five Galactic YOCs and one SFR that are included in the \textit{Gaia}-ESO survey.

In particular, for a young cluster star an EW analysis that only uses Fe\,{\sc i} lines returns a value for the $\xi$ parameter that is too high, as shown by the model lines, which are too strong compared to the observed lines. This indicates that the derived Fe line abundances depend on the optical depth of line formation, as suggested by \cite{reddy}. We also confirm that this effect is weak in old stars, as we showed for the Gaia benchmark stars, for which we obtain the same results as \cite{2015jofre,2018jofre} and \textit{Gaia}-ESO. 

Our method consists of a combination of Fe and Ti lines to derive $T\rm{_{eff}}$  by zeroing the trend between the individual line abundances and the E.P. For $\log g$ and the $\xi$ parameter, we only use\textit{} Ti lines to avoid possible complications due to the use of Fe lines, because the Fe lines form in a wider range of optical depth and the strongest lines can be affected by the hot active chromosphere. The comparison with \textit{Gaia}-ESO iDR5 results showed that while for $T\rm{_{eff}}$ and $\log g$ we obtain comparable measurements, a most dramatic effect is seen for $\xi$. Overall, we note an overestimation of this parameter, with the largest differences seen for clusters younger than 100\,Myr. 

We plot the metallicity distribution as a function of the open cluster age in Fig.~\ref{age_met}, where the empty circles represent the clusters in \cite{neto} for which high-quality determination of metallicity are available \citep{heiter} and with $7.5<R\rm{_{gal}}<9$\,kpc. The coloured stars represent the clusters we have analysed with our new determination of [Fe/H]. The empty stars represent our clusters with the metallicity derived with the standard analysis. The blue triangles represent the \textit{Gaia}-ESO clusters analysed in \cite{magrini} (intermediate-age and old clusters), for which we take the median [Fe/H] value reported in the same paper. In the case of NGC\,2264, we take the age estimate by \cite{venuti} (in contrast with the estimate by \citealt{neto}, who report 10$\pm$10\,Myr), both for our new metallicity estimate and the value obtained with the standard analysis.  The grey crosses, instead, represent the field stars taken from \cite{ben}, from which we select only thin-disc stars, those with a probability ratio TD/D$<$0.5, where TD is the probability of being a thick-disc star, and D corresponds to the probability of being a thin-disc star. A more detailed description of these parameters can be found in the Appendix A in \cite{ben}.  Also, we exclude those field stars with a difference between the upper and lower age limit that is larger than 4\,Gyr, in order to exclude those stars with large age uncertainties. 
When we consider our new estimates of the abundances and the model for the solar surroundings developed by \cite{minchev}, no peculiar chemical evolution of the Galaxy seems required. The model of \cite{minchev} does not extend to ages younger than $\sim$ 60\,Myr, as already noted by \cite{Spina17}. However, we expect an enrichment of nearly 0.10-0.15\,dex at $R\rm{_{Gal}}$ of 7.5-9\,kpc in the last 4-5\,Gyr. We might also expect the model to extend to the present time with a flat, continuous behaviour, but very likely not towards sub-solar metallicities, as the standard analysis seems to suggest instead. All of our new estimates of [Fe/H], ranging from 0.04 to 0.12\,dex, lie within the predictions of the Galactic chemical evolution model. We also note that the sample of \cite{neto} lies lower than predicted by the theoretical model. We do not have a conclusive explanation for this behaviour: it might be the combination of different factors, for example, the use of the standard analysis, but also the fact that the metallicity determinations in \cite{neto} are a combination of different studies.

We also analysed the effect of stellar activity. In Sec.\ref{S5} we reported that the difference between EWs measured in a young (30\,Myr) solar analogue and the Sun is larger for lines that form in the outer layers of the atmosphere. Fig.\ref{ew_tau} showed that lines forming at $\log \tau_{5000}<-2.5$ are too strong in the young star compared with the values measured in the Sun. This might cause higher values of the $\xi$ parameter when it is derived based on the Fe lines. We also find a dependence of the $\xi$ values on the chromospheric activity index $\log\,\rm{R^{\prime}_{HK}}$, which is stronger for the GESiDR5 results than for the values we derived with the new approach. This also confirms that for the Gaia benchmarck stars, which are quiet stars, we obtain the same values with both methods. 

Finally, the revised metallicities might also affect the isochrone-derived ages of very young stars. Redder or cooler stars mimic younger ages if a higher metallicity is assumed. The isochrone-based ages may therefore be slightly different when this is performed for stars whose colour-magnitude diagram depends on [Fe/H].

\begin{table*}
\centering
\caption{Mean abundances and abundance ratios of each cluster.}   
\label{mean_Ab}      
\centering  
\setlength\tabcolsep{10pt}
\small
\begin{tabular}{lccccccr}    
\toprule       
 & \bf{IC\,2391} & \bf{IC\,2602} & \bf{IC\,4665}  & \bf{NGC\,2264}$^{\ast}$ & \bf{NGC\,2516} & \bf{NGC\,2547}\\ 
\midrule
$\rm{[Fe/H]}$ & 0.04$\pm$0.01 & 0.04$\pm$0.01 & 0.12$\pm$0.02 &  0.11$\pm$0.02 & 0.08$\pm$0.01 & 0.10$\pm$0.01 \\
$\rm{[Ti/H]}$ & 0.02$\pm$0.01 & 0.03$\pm$0.01 & 0.11$\pm$0.03 &  0.19$\pm$0.09 & 0.09$\pm$0.01  & 0.10$\pm$0.01  \\
$\rm{[Na/Fe]}$ & $-$0.03$\pm$0.02 & 0.02$\pm$0.01 & 0.04$\pm$0.03 & 0.06$\pm$0.04 & $-$0.03$\pm$0.01 & $-$0.005$\pm$0.004 \\
$\rm{[Mg/Fe]}$ & 0.02$\pm$0.02 & $-$0.03$\pm$0.01 & - & $-$0.02$\pm$0.04 & $-$0.07$\pm$0.02 & 0.01$\pm$0.03 \\
$\rm{[Al/Fe]}$ & 0.03$\pm$0.02 & $-$0.01$\pm$0.03 & - & - & 0.00$\pm$0.04 & 0.04$\pm$0.04 \\
$\rm{[Si/Fe]}$ & 0.00$\pm$0.01 & 0.01$\pm$0.03 & 0.03$\pm$0.02 & $-$0.02$\pm$0.05 & $-$0.02$\pm$0.01 & $-$0.01$\pm$0.02 \\
$\rm{[Ca/Fe]}$ & 0.075$\pm$0.004 & 0.08$\pm$0.02 & 0.07$\pm$0.02 & 0.02$\pm$0.04 & 0.04$\pm$0.01 & 0.07$\pm$0.01 \\
$\rm{[Ti/Fe]}$ &  0.01$\pm$0.01 & 0.01$\pm$0.02 & 0.00$\pm$0.04 & 0.03$\pm$0.06 & 0.008$\pm$0.004 & 0.01$\pm$0.02 \\
$\rm{[Ti/Fe]_{II}}$ & $-$0.05$\pm$0.01 & $-$0.02$\pm$0.01 & $-$0.02$\pm$0.04 & 0.04$\pm$0.05 & $-$0.01$\pm$0.01 & $-$0.03$\pm$0.01 \\
$\rm{[Cr/Fe]}$ & 0.01$\pm$0.03 & 0.05$\pm$0.01 & 0.08$\pm$0.02 & 0.02$\pm$0.06 & 0.02$\pm$0.01 & 0.07$\pm$0.04 \\
$\rm{[Ni/Fe]}$ & $-$0.04$\pm$0.02 & 0.023$\pm$0.003 & 0.06$\pm$0.01 & 0.00$\pm$0.02 & 0.03$\pm$0.01 & 0.03$\pm$0.02 \\
\hline               
\end{tabular}
\begin{tablenotes}
\small
\item $^{\ast}$ For this cluster we analysed only one star, therefore the uncertainty is the quadratic sum of $\sigma_1$ and $\sigma_2$.
\end{tablenotes}
\end{table*}

\begin{acknowledgements}
We thank the anonymous referee for very helpful comments and suggestions.
This work is based on data products from observations made with ESO Telescopes at the La Silla Paranal Observatory under programme ID 188.B-3002. These data products have been processed by the Cambridge Astronomy Survey Unit (CASU) at the Institute of Astronomy, University of Cambridge, and by the FLAMES/UVES reduction team at INAF/Osservatorio Astrofisico di Arcetri. These data have been obtained from the Gaia-ESO Survey Data Archive, prepared and hosted by the Wide Field Astronomy Unit, Institute for Astronomy, University of Edinburgh, which is funded by the UK Science and Technology Facilities Council.

This work was partly supported by the European Union FP7 programme through ERC grant number 320360 and by the Leverhulme Trust through grant RPG-2012-541. We acknowledge the support from INAF and Ministero dell' Istruzione, dell' Universit\`a' e della Ricerca (MIUR) in the form of the grant "Premiale VLT 2012". The results presented here benefit from discussions held during the Gaia-ESO workshops and conferences supported by the ESF (European Science Foundation) through the GREAT Research Network Programme.

V.A. is supported by FCT - Fundação para a Ciência e a Tecnologia through national funds and by FEDER through COMPETE2020 - Programa Operacional Competitividade e Internacionalização by these grants: Investigador FCT contract nr. IF/00650/2015/CP1273/CT0001; UID/FIS/04434/2019; PTDC/FIS-AST/28953/2017 \& POCI-01-0145-FEDER-028953 and PTDC/FIS-AST/32113/2017 \& POCI-01-0145-FEDER-032113.

F.J.E. acknowledges financial support from the Spanish MINECO/FEDER through grant AyA2017-84089.

T.B. was supported by the project grant ’The New Milky Way’ from the Knut and Alice Wallenberg Foundation.
U.H. acknowledges support from the Swedish National Space Agency (SNSA/Rymdstyrelsen).
S.G.S acknowledges the support by Fundação para a Ciência e Tecnologia
(FCT) through national funds and a research grant (project ref.
UID/FIS/04434/2013, and PTDC/FIS-AST/7073/2014). S.G.S. also acknowledge
the support from FCT through Investigador FCT contract of reference
IF/00028/2014 and POPH/FSE (EC) by FEDER funding through the program -- Programa Operacional de Factores de Competitividade -- COMPETE

\end{acknowledgements}

\bibliographystyle{aa} 
\bibliography{baratella_GES} 

\begin{appendix}
\section{Additional tables}

\begin{table*}[]
\caption{Different values of $T\rm{_{eff}}$ derived with photometry, Fe lines alone, and Fe and Ti simultaneously. We also report the number of the Fe and Ti lines that we measured for each star. In the last column, we repot $v\,{\rm sin}\, i$ values taken from \textit{Gaia}-ESO iDR5.}    
\label{TabTs}      
\centering  
\setlength\tabcolsep{15pt}
\small
\begin{tabular}{lccccccr }    
\toprule       
CNAME & $T_{\rm{eff,phot}}$& $T\rm{_{Fe\,I}}$ &  $T\rm{_{Fe\,I+Ti\,I}}$ & n$\rm_{Fe\,I}$ & n$\rm_{Ti\,I}$ &  $v\,{\rm sin}\, i$\\ 
 & \footnotesize{(K)} & \footnotesize{(K)}&\footnotesize{(K)} &  & & \footnotesize{~km\, s$^{-1}$}\\
\midrule
&&\bf{IC\,2391}\\

08365498$-$5308342& 5215$\pm$118 & 5150$\pm$100 &  5150$\pm$100 & 39 & 17 & 9.88 \\
08440521$-$5253171& 5471$\pm$103 & 5471$\pm$50 &  5471$\pm$75 & 35 & 9 & 19.47\\
\\
&&\bf{IC\,2602}\\

10440681$-$6359351& 5600$\pm$162 & 5500$\pm$100 &  5500$\pm$75 & 32 & 10 & 12.87\\
10442256$-$6415301 & 5825$\pm$117 & 5765$\pm$75 &  5765$\pm$75 & 40 & 12 & 11.54 \\
10481856$-$6409537 & 5753$\pm$114 & 5700$\pm$100 &  5680$\pm$100 & 29 & 8 & 13.81\\
\\
&&\bf{IC\,4665}\\

17442711+0547196 & 5397$\pm$103 &  5280$\pm$75 &  5397$\pm$75 & 15 & 9 & 14.92 \\
17445810+0551329 & 5650$\pm$118 & 5600$\pm$75 &  5575$\pm$75 &  31 & 11 & 10.72  \\
17452508+0551388 &5321$\pm$108 & 5200$\pm$100 &  5271$\pm$100 & 17 & 10 & 14.13 \\
\\
&&\bf{NGC\,2264}\\

06405694+0948407 & 6081$\pm$162 & 6150$\pm$75 &  6150$\pm$75  & 30 & 7  & 20.02\\
\\
&&\bf{NGC\,2516}\\

07544342$-$6024437 & 5487$\pm$133 & 5325$\pm$75 &  5300$\pm$100  & 46 & 16 & 6.51\\
07550592$-$6104294 & 5570$\pm$107 & 5500$\pm$75 & 5500$\pm$75 &  30 & 10 & 11.64 \\
07551977$-$6104200 &6064$\pm$161 & 6064$\pm$100 &  6050$\pm$100  & 33 & 8 & 14.24\\
07553236$-$6023094 & 5739$\pm$114 & 5650$\pm$75 &  5625$\pm$75 & 33 & 10 & 10.21 \\
07564410$-$6034523 & 5708$\pm$112 & 5600$\pm$75 &  5600$\pm$75 & 28 & 10 & 9.43\\
07573608$-$6048128 & 5592$\pm$108 & 5572$\pm$100 &  5572$\pm$75 & 35 & 12 & 7.50  \\
07574792$-$6056131 & 5617$\pm$109 & 5525$\pm$75 &  5515$\pm$75  & 35 & 18 &6.54 \\
07575215$-$6100318 & 5287$\pm$97 & 5200$\pm$75 &  5170$\pm$75  & 33 & 10 & 7.76\\
07583485$-$6103121 & 5708$\pm$112 & 5758$\pm$75 &  5730$\pm$75 & 28 & 10 & 12.04\\
07584257$-$6040199 &5643$\pm$110 & 5525$\pm$75 &  5500$\pm$75 & 33 & 10 & 10.17\\
08000944$-$6033355 &5753$\pm$145 & 5700$\pm$75 &  5700$\pm$75 & 32 & 11 & 8.90\\
08013658$-$6059021 & 5673$\pm$111 & 5600$\pm$75 & 5575$\pm$75 & 28 & 9 & 10.08 \\
\\
&&\bf{NGC\,2547}\\

08102854$-$4856518 & 5800$\pm$148 & 5800$\pm$100 & 5800$\pm$100 & 24 & 6 & 18.06\\
08110139$-$4900089 & 5453$\pm$103 & 5250$\pm$75 & 5353$\pm$100& 32 & 12 & 9.93\\
\hline                  
\end{tabular}
\end{table*}

\begin{sidewaystable*}
\renewcommand\arraystretch{1.0}
\caption{Derived stellar parameters and abundances with the final models and comparison with the photometric values. }
\setlength\tabcolsep{7pt}
\small
\label{abb}
\centering
\begin{tabular}{lccccccccccccr}
\toprule
CNAME  & log\,$g_{phot}$ & $\xi_{phot}$ & $T\rm{_{spec}}$ & log\,$g_{spec}$ & $\xi_{spec}$ & [Fe/H]$_{\rm{I}}\pm\sigma_{1}\pm\sigma_{2}$  &[Fe/H]$_{\rm{II}}\pm\sigma_{1}\pm\sigma_{2}$ &  [Ti/H]$_{\rm{I}}\pm\sigma_{1}\pm\sigma_{2}$ &  [Ti/H]$_{\rm{II}}\pm\sigma_{1}\pm\sigma_{2}$  \\
 & \footnotesize{(dex)} & \footnotesize{(km\,s$^{-1}$)}&\footnotesize{(K)} & \footnotesize{(dex)}&  \footnotesize{(km\,s$^{-1}$)} & \\
\midrule

& & & & &&\bf{IC\,2391}\\
08365498-5308342 & 4.47$\pm$0.06 & 0.70$\pm$0.03 & 5215$\pm$100 & 4.35$\pm$0.10 & 0.85$\pm$0.10 & 0.00$\pm$0.01$\pm$0.06 & 0.09$\pm$0.03$\pm$0.08 & 0.01$\pm$0.01$\pm$0.13 & 0.02$\pm$0.03$\pm$0.05 & \\
08440521-5253171 & 4.28$\pm$0.06 & 0.88$\pm$0.04 & 5471$\pm$100 & 4.20$\pm$0.10 & - & 0.00$\pm$0.02$\pm$0.07 & 0.05$\pm$0.03$\pm$0.06 & 0.01$\pm$0.02$\pm$0.12 & 0.02$\pm$0.04$\pm$0.06 \\
\\
&&&&&&\bf{IC\,2602}\\
10440681-6359351 & 4.46$\pm$0.07 & 0.92$\pm$0.07 & 5525$\pm$75 & 4.38$\pm$0.15 & 1.00$\pm$0.20 & 0.00$\pm$0.01$\pm$0.06 & 0.08$\pm$0.03$\pm$0.06 & 0.06$\pm$0.02$\pm$0.10 & 0.07$\pm$0.04$\pm$0.05\\
10442256-6415301 & 4.49$\pm$0.07 & 1.04$\pm$0.06 & 5775$\pm$75 & 4.49$\pm$0.10 & 1.15$\pm$0.10 & 0.04$\pm$0.01$\pm$0.07 & 0.05$\pm$0.02$\pm$0.05 & 0.02$\pm$0.02$\pm$0.10 & 0.02$\pm$0.03$\pm$0.04\\
10481856-6409537 & 4.18$\pm$0.06 & 1.09$\pm$0.05 & 5680$\pm$100 & 4.10$\pm$0.10 & - & 0.03$\pm$0.02$\pm$0.07 & 0.06$\pm$0.03$\pm$0.06 & 0.01$\pm$0.02$\pm$0.10 & 0.02$\pm$0.04$\pm$0.06\\
\\
&&&&&&\bf{IC\,4665}\\
17442711+0547196 & 4.43$\pm$0.03 & 0.80$\pm$0.04 & 5380$\pm$75 & 4.48$\pm$0.10 & - & 0.14$\pm$0.02$\pm$0.05 & 0.17$\pm$0.02$\pm$0.06 & 0.05$\pm$0.03$\pm$0.09 & 0.04$\pm$0.02$\pm$0.04\\
17445810+0551329 & 4.49$\pm$0.03 & 0.91$\pm$0.06 & 5575$\pm$75 & 4.47$\pm$0.10 & 0.96$\pm$0.10 & 0.12$\pm$0.01$\pm$0.07 & 0.13$\pm$0.04$\pm$0.04 & 0.13$\pm$0.02$\pm$0.11 & 0.15$\pm$0.03$\pm$0.05\\
17452508+0551388 & 4.48$\pm$0.03 & 0.75$\pm$0.04 & 5300$\pm$100 & 4.27$\pm$0.15 & 1.03$\pm$0.10 & 0.05$\pm$0.03$\pm$0.06 & 0.11$\pm$0.03$\pm$0.08 & 0.12$\pm$0.02$\pm$0.13 & 0.14$\pm$0.04$\pm$0.06\\
\\
&&&&&&\bf{NGC\,2264}\\
06405694+0948407 & 4.15$\pm$0.11 & 1.29$\pm$0.08 & 6150$\pm$75 & 4.05$\pm$0.10 & - & 0.10$\pm$0.02$\pm$0.07 & 0.11$\pm$0.02$\pm$0.05 & 0.13$\pm$0.02$\pm$0.09 & 0.15$\pm$0.04$\pm$0.05\\
\\
&&&&&&\bf{NGC\,2516}\\
07544342-6024437 & 4.69$\pm$0.06 & 0.78$\pm$0.06 & 5430$\pm$100 & 4.51$\pm$0.10 & 1.05$\pm$0.10 & 0.03$\pm$0.01$\pm$0.07 & 0.04$\pm$0.03$\pm$0.08 & 0.02$\pm$0.02$\pm$0.11 & 0.03$\pm$0.02$\pm$0.05\\
07550592-6104294 & 4.51$\pm$0.05 & 0.87$\pm$0.04 & 5550$\pm$75 & 4.20$\pm$0.10 & 0.95$\pm$0.10 & 0.05$\pm$0.03$\pm$0.07 & 0.08$\pm$0.02$\pm$0.06 & 0.04$\pm$0.02$\pm$0.09 & 0.04$\pm$0.02$\pm$0.05\\
07551977-6104200 & 4.59$\pm$0.06 & 1.15$\pm$0.10 & 6050$\pm$100 & 4.62$\pm$0.10 & - & 0.12$\pm$0.02$\pm$0.06 & 0.12$\pm$0.03$\pm$0.05 & 0.12$\pm$0.02$\pm$0.08 & 0.12$\pm$0.01$\pm$0.05\\
07553236-6023094 & 4.59$\pm$0.05 & 0.94$\pm$0.05 & 5700$\pm$75 & 4.52$\pm$0.10 & 1.15$\pm$0.15 & 0.09$\pm$0.01$\pm$0.07 & 0.13$\pm$0.01$\pm$0.07 & 0.12$\pm$0.02$\pm$0.08 & 0.13$\pm$0.02$\pm$0.05\\
07564410-6034523 & 4.61$\pm$0.05 & 0.92$\pm$0.05 & 5650$\pm$75 & 4.45$\pm$0.10 & 1.02$\pm$0.10 & 0.13$\pm$0.02$\pm$0.06 & 0.14$\pm$0.04$\pm$0.05 & 0.13$\pm$0.02$\pm$0.08 & 0.13$\pm$0.03$\pm$0.04\\
07573608-6048128 & 4.62$\pm$0.05 & 0.85$\pm$0.05 & 5625$\pm$100 & 4.55$\pm$0.10 & 1.03$\pm$0.10 & 0.11$\pm$0.01$\pm$0.07 & 0.10$\pm$0.02$\pm$0.08 & 0.11$\pm$0.03$\pm$0.11 & 0.11$\pm$0.04$\pm$0.05\\
07574792-6056131 & 4.73$\pm$0.05 & 0.84$\pm$0.05 & 5580$\pm$75 & 4.57$\pm$0.10 & 1.08$\pm$0.10 & 0.08$\pm$0.01$\pm$0.07 & 0.10$\pm$0.03$\pm$0.07 & 0.10$\pm$0.02$\pm$0.09 & 0.10$\pm$0.04$\pm$0.05\\
07575215-6100318 & 4.76$\pm$0.05 & 0.67$\pm$0.03 & 5275$\pm$100 & 4.54$\pm$0.10 & 0.98$\pm$0.10 & 0.07$\pm$0.02$\pm$0.05 & 0.10$\pm$0.05$\pm$0.08 & 0.10$\pm$0.02$\pm$0.09 & 0.11$\pm$0.02$\pm$0.05\\
07583485-6103121 & 4.45$\pm$0.05 & 0.96$\pm$0.05 & 5758$\pm$100 & 4.43$\pm$0.10 & 1.05$\pm$0.15 & 0.02$\pm$0.02$\pm$0.06 & 0.04$\pm$0.03$\pm$0.03 & 0.03$\pm$0.03$\pm$0.08 & 0.03$\pm$0.03$\pm$0.04\\
07584257-6040199 & 4.68$\pm$0.05 & 0.86$\pm$0.05 & 5550$\pm$75 & 4.48$\pm$0.10 & 0.98$\pm$0.10 & 0.09$\pm$0.01$\pm$0.06 & 0.08$\pm$0.04$\pm$0.06 & 0.09$\pm$0.02$\pm$0.09 & 0.09$\pm$0.02$\pm$0.05\\
08000944-6033355 & 4.60$\pm$0.06 & 0.95$\pm$0.07 & 5675$\pm$100 & 4.38$\pm$0.10 & 0.90$\pm$0.10 & 0.05$\pm$0.02$\pm$0.05 & 0.06$\pm$0.02$\pm$0.05 & 0.06$\pm$0.02$\pm$0.08 & 0.08$\pm$0.01$\pm$0.04\\
08013658-6059021 & 4.54$\pm$0.06 & 0.92$\pm$0.05 & 5585$\pm$100 & 4.32$\pm$0.10 & - & 0.06$\pm$0.02$\pm$0.06 & 0.07$\pm$0.02$\pm$0.05 & 0.08$\pm$0.02$\pm$0.09 & 0.08$\pm$0.02$\pm$0.04\\
\\
&&&&&&\bf{NGC\,2547}\\
08102854-4856518 & 4.34$\pm$0.05 & 1.04$\pm$0.07 & 5800$\pm$100 & 4.20$\pm$0.10 & - & 0.11$\pm$0.02$\pm$0.07 & 0.14$\pm$0.03$\pm$0.05 & 0.10$\pm$0.01$\pm$0.10 & 0.11$\pm$0.03$\pm$0.04\\
08110139-4900089 & 4.51$\pm$0.05 & 0.80$\pm$0.04 & 5375$\pm$100 & 4.50$\pm$0.10 & 1.05$\pm$0.10 & 0.05$\pm$0.01$\pm$0.06 & 0.10$\pm$0.04$\pm$0.08 & 0.08$\pm$0.02$\pm$0.12 & 0.09$\pm$0.03$\pm$0.05\\
\\

\bottomrule
\end{tabular}

\end{sidewaystable*}

\begin{sidewaystable*}
\renewcommand\arraystretch{1.0}
\caption{Abundance ratios for different species. }
\setlength\tabcolsep{1.0pt}
\small
\label{ratios}
\centering
\begin{tabular}{lcccccccccccccccr}
\toprule
CNAME& [Na/Fe]$\pm\sigma_1\pm\sigma_2$ & [Mg/Fe]$\pm\sigma_1\pm\sigma_2$ & [Al/Fe]$\pm\sigma_1\pm\sigma_2$ & [Si/Fe]$\pm\sigma_1\pm\sigma_2$ & [Ca/Fe]$\pm\sigma_1\pm\sigma_2$ & [Ti/Fe]$_{\rm{I}}\pm\sigma_1\pm\sigma_2$ & [Ti/Fe]$_{\rm{II}}$ $\pm\sigma_1\pm\sigma_2$ & [Cr/Fe]$\pm\sigma_1\pm\sigma_2$ & [Ni/Fe]$\pm\sigma_1\pm\sigma_2$\\
\\
\midrule
&&&&&\bf{IC\,2391}\\
08365498-5308342& 0.00$\pm$0.01$\pm$0.04 & 0.05$\pm$0.01$\pm$0.06 & 0.03$\pm$0.01$\pm$0.03 & $-$0.01$\pm$0.02$\pm$0.07 & 0.08$\pm$0.01$\pm$0.09 & 0.01$\pm$0.01$\pm$0.07 & $-$0.07$\pm$0.04$\pm$0.10 & $-$0.04$\pm$0.04$\pm$0.14 & $-$0.06$\pm$0.02$\pm$0.10\\
08440521-5253171 & $-$0.06$\pm$0.02$\pm$0.13 & $-$0.01$\pm$0.02$\pm$0.03 & - & 0.01$\pm$0.02$\pm$0.08 & 0.07$\pm$0.02$\pm$0.09 & 0.01$\pm$0.02$\pm$0.04 & $-$0.03$\pm$0.03$\pm$0.08 & 0.05$\pm$0.02$\pm$0.03 & $-$0.01$\pm$0.02$\pm$0.02\\
\\
&&&&&\bf{IC\,2602}\\
10440681-6359351 & 0.02$\pm$0.06$\pm$0.10 & $-$0.01$\pm$0.02$\pm$0.17 & - & 0.08$\pm$0.02$\pm$0.13 & 0.12$\pm$0.03$\pm$0.14 & 0.06$\pm$0.02$\pm$0.11 & 0.00$\pm$0.04$\pm$0.03 & 0.05$\pm$0.03$\pm$0.15 & 0.02$\pm$0.03$\pm$0.16\\
10442256-6415301  & 0.02$\pm$0.01$\pm$0.02 & $-$0.05$\pm$0.01$\pm$0.06 & - & $-$0.03$\pm$0.02$\pm$0.07 & 0.04$\pm$0.02$\pm$0.07 & $-$0.02$\pm$0.02$\pm$0.07 & $-$0.03$\pm$0.03$\pm$0.06 & 0.08$\pm$0.03$\pm$0.14 & 0.02$\pm$0.01$\pm$0.03\\
10481856-6409537 & 0.02$\pm$0.01$\pm$0.04 & $-$0.03$\pm$0.01$\pm$0.05 & $-$0.01$\pm$0.01$\pm$0.05 & $-$0.02$\pm$0.04$\pm$0.08 & 0.07$\pm$0.05$\pm$0.10 & $-$0.02$\pm$0.02$\pm$0.04 & $-$0.03$\pm$0.04$\pm$0.06 & 0.03$\pm$0.05$\pm$0.04 & 0.03$\pm$0.02$\pm$0.06\\
\\
&&&&&\bf{IC\,4665}\\
17442711+0547196 & $-$0.02$\pm$0.02$\pm$0.14 & - & - & 0.01$\pm$0.02$\pm$0.06 & 0.03$\pm$0.04$\pm$0.02 & $-$0.09$\pm$0.03$\pm$0.16 & $-$0.12$\pm$0.02$\pm$0.14 & 0.12$\pm$0.02$\pm$0.14 & 0.04$\pm$0.02$\pm$0.15\\
17445810+0551329  & 0.05$\pm$0.01$\pm$0.02 & - & - & $-$0.01$\pm$0.02$\pm$0.08 & 0.09$\pm$0.03$\pm$0.04 & 0.02$\pm$0.02$\pm$0.04 & 0.02$\pm$0.03$\pm$0.04 & 0.06$\pm$0.02$\pm$0.10 & 0.07$\pm$0.02$\pm$0.06\\
17452508+0551388  & 0.10$\pm$0.03$\pm$0.16 & - & - & 0.08$\pm$0.03$\pm$0.17 & 0.09$\pm$0.03$\pm$0.04 & 0.08$\pm$0.03$\pm$0.12 & 0.05$\pm$0.04$\pm$0.08 & 0.06$\pm$0.03$\pm$0.04 & 0.07$\pm$0.02$\pm$0.14\\

\\
&&&&&\bf{NGC\,2264}\\
06405694+0948407 & 0.06$\pm$0.02$\pm$0.07 & $-$0.02$\pm$0.02$\pm$0.06 & - & $-$0.01$\pm$0.02$\pm$0.04 & 0.02$\pm$0.01$\pm$0.04 & 0.03$\pm$0.02$\pm$0.06 & 0.04$\pm$0.03$\pm$0.14 & 0.02$\pm$0.03$\pm$0.04 & 0.00$\pm$0.02$\pm$0.02\\
\\
&&&&&\bf{NGC\,2516}\\
07544342-6024437 & $-$0.03$\pm$0.01$\pm$0.13 & 0.06$\pm$0.01$\pm$0.04 & 0.06$\pm$0.03$\pm$0.02 & 0.00$\pm$0.02$\pm$0.11 & 0.04$\pm$0.02$\pm$0.02 & 0.00$\pm$0.01$\pm$0.07 & $-$0.01$\pm$0.03$\pm$0.14 & 0.04$\pm$0.04$\pm$0.05 & 0.02$\pm$0.02$\pm$0.03\\
07550592-6104294& $-$0.07$\pm$0.02$\pm$0.14 & $-$0.11$\pm$0.01$\pm$0.12 & - & $-$0.02$\pm$0.03$\pm$0.06 & 0.05$\pm$0.04$\pm$0.10 & $-$0.02$\pm$0.02$\pm$0.04 & $-$0.07$\pm$0.02$\pm$0.11 & 0.04$\pm$0.04$\pm$0.03 & 0.01$\pm$0.03$\pm$0.04\\
07551977-6104200  & 0.03$\pm$0.01$\pm$0.04 & $-$0.13$\pm$0.01$\pm$0.13 & - & $-$0.03$\pm$0.02$\pm$0.06 & 0.06$\pm$0.01$\pm$0.09 & 0.00$\pm$0.02$\pm$0.06 & $-$0.01$\pm$0.03$\pm$0.06 & 0.07$\pm$0.02$\pm$0.06 & 0.05$\pm$0.03$\pm$0.09\\
07553236-6023094 & $-$0.05$\pm$0.03$\pm$0.04 & $-$0.10$\pm$0.01$\pm$0.15 & - & $-$0.03$\pm$0.03$\pm$0.09 & 0.06$\pm$0.02$\pm$0.07 & 0.03$\pm$0.02$\pm$0.03 & $-$0.01$\pm$0.02$\pm$0.05 & $-$0.02$\pm$0.01$\pm$0.07 & 0.02$\pm$0.02$\pm$0.03\\
07564410-6034523  & $-$0.01$\pm$0.01$\pm$0.15 & $-$0.12$\pm$0.01$\pm$0.18 & - & $-$0.06$\pm$0.03$\pm$0.12 & 0.05$\pm$0.04$\pm$0.010 & 0.00$\pm$0.01$\pm$0.03 & $-$0.01$\pm$0.01$\pm$0.04 & 0.02$\pm$0.04$\pm$0.02 & 0.00$\pm$0.02$\pm$0.06\\
07573608-6048128  & 0.00$\pm$0.02$\pm$0.04 & $-$0.10$\pm$0.01$\pm$0.16 & - & $-$0.04$\pm$0.01$\pm$0.10 & 0.04$\pm$0.02$\pm$0.03 & 0.00$\pm$0.02$\pm$0.04 & 0.02$\pm$0.03$\pm$0.10 & $-$0.02$\pm$0.02$\pm$0.04 & $-$0.01$\pm$0.02$\pm$0.06\\
07574792-6056131   & $-$0.09$\pm$0.01$\pm$0.1 & - & $-$0.06$\pm$0.01$\pm$0.14 & $-$0.05$\pm$0.02$\pm$0.08 & 0.00$\pm$0.02$\pm$0.10 & 0.02$\pm$0.02$\pm$0.06 & 0.01$\pm$0.04$\pm$0.03 & 0.02$\pm$0.02$\pm$0.09 & 0.03$\pm$0.02$\pm$0.05\\
07575215-6100318  & 0.03$\pm$0.05$\pm$0.05 & - & - & $-$0.04$\pm$0.01$\pm$0.11 & 0.02$\pm$0.03$\pm$0.07 & 0.03$\pm$0.02$\pm$0.04 & 0.00$\pm$0.05$\pm$0.05 & 0.05$\pm$0.04$\pm$0.02 & 0.03$\pm$0.02$\pm$0.03\\
07583485-6103121 & $-$0.04$\pm$0.02$\pm$0.09 & $-$0.05$\pm$0.02$\pm$0.09 & - & 0.03$\pm$0.02$\pm$0.08 & 0.08$\pm$0.02$\pm$0.09 & $-$0.01$\pm$0.02$\pm$0.03 & $-$0.02$\pm$0.03$\pm$0.02 & $-$0.06$\pm$0.04$\pm$0.10 & 0.06$\pm$0.02$\pm$0.12\\
07584257-6040199 & 0.00$\pm$0.01$\pm$0.05 & $-$0.09$\pm$0.01$\pm$0.11 & - & $-$0.03$\pm$0.03$\pm$0.07 & 0.03$\pm$0.03$\pm$0.05 & 0.01$\pm$0.02$\pm$0.03 & 0.01$\pm$0.02$\pm$0.03 & 0.08$\pm$0.01$\pm$0.09 & 0.01$\pm$0.02$\pm$0.03\\
08000944-6033355  & $-$0.07$\pm$0.04$\pm$0.02 & $-$0.05$\pm$0.01$\pm$0.02 & - & $-$0.03$\pm$0.05$\pm$0.05 & 0.03$\pm$0.03$\pm$0.01 & 0.01$\pm$0.02$\pm$0.03 & 0.02$\pm$0.02$\pm$0.03 & 0.04$\pm$0.05$\pm$0.03 & 0.05$\pm$0.02$\pm$0.01\\
08013658-6059021  & $-$0.01$\pm$0.01$\pm$0.15 & 0.00$\pm$0.01$\pm$0.09 & - & 0.02$\pm$0.02$\pm$0.07 & 0.06$\pm$0.04$\pm$0.09 & 0.02$\pm$0.02$\pm$0.04 & 0.00$\pm$0.02$\pm$0.03 & - & 0.07$\pm$0.01$\pm$0.10\\
\\
&&&&&\bf{NGC\,2547}\\
08102854-4856518  & $-$0.01$\pm$0.04$\pm$0.03 & 0.01$\pm$0.02$\pm$0.08 & - & 0.02$\pm$0.03$\pm$0.08 & 0.05$\pm$0.02$\pm$0.09 & $-$0.02$\pm$0.01$\pm$0.05 & $-$0.04$\pm$0.03$\pm$0.06 & - & 0.05$\pm$0.04$\pm$0.09\\
08110139-4900089  & 0.00$\pm$0.01$\pm$0.03 & - & 0.04$\pm$0.01$\pm$0.02 & $-$0.03$\pm$0.02$\pm$0.07 & 0.09$\pm$0.02$\pm$0.13 & 0.04$\pm$0.02$\pm$0.07 & $-$0.01$\pm$0.04$\pm$0.07 & 0.07$\pm$0.05$\pm$0.04 & 0.01$\pm$0.02$\pm$0.03\\
\bottomrule
\end{tabular}
\end{sidewaystable*}

\newpage

\end{appendix}


\end{document}